\documentclass[preprint, nofootinbib, tightenlines, eqsecnum, 
floats, aps,amsfonts]{revtex4}

\usepackage{amsmath,graphics,color}

\newcommand  {\Rbar} {{\mbox{\rm$\mbox{I}\!\mbox{R}$}}}

\newcommand{\Lie}[0]{{\cal L}\, }

\newcommand{\vS}{\widetilde{\mbox{\boldmath$\epsilon$}}}
\newcommand{\tq}{\tilde{q}}
\def\l{\ell}
\newcommand{\tl}{\theta_{(\ell)}}
 \newcommand{\tn}{\theta_{(n)}}
\newcommand{\cV}{\mathcal{V}}
\newcommand{\tom}{\tilde{\omega}}
\newcommand{\norm}{| \mspace{-2mu} |}

\newcommand{\nn}{\nonumber}
\newcommand{\be}{\begin{equation}}
\newcommand{\ee}{\end{equation}}
\newcommand{\bea}{\begin{eqnarray}}
\newcommand{\eea}{\end{eqnarray}}

\begin{document}

\title{Isolated, slowly evolving, and dynamical trapping horizons: \\ 
geometry and mechanics from surface deformations}
\author {Ivan Booth\footnote{E-mail:
ibooth@math.mun.ca}}
\affiliation{
Department of Mathematics and Statistics, Memorial University of
Newfoundland \\ 
St. John's, Newfoundland and Labrador, A1C 5S7, Canada}
\author{Stephen Fairhurst\footnote{E-mail:
sfairhur@gravity.phys.uwm.edu, Current Address: California Institute 
of
Technology}}
\affiliation{Department of Physics, University of Wisconsin--Milwaukee 
\\  
Milwaukee, Wisconsin, 53201, USA}

%
\begin{abstract}
%

We study the geometry and dynamics of both isolated and dynamical
trapping horizons by considering the allowed variations of their
foliating two-surfaces. This provides a common framework that may be
used to consider both their possible evolutions and their deformations
as well as derive the well-known flux laws. Using this framework, we
unify much of what  is already known about these objects as well as
derive some new results. In particular we characterize and study the
``almost-isolated" trapping horizons known as slowly evolving 
horizons.
It is for these horizons that a dynamical first law holds and this is
analogous and  closely related to the Hawking-Hartle formula for event
horizons.

\end{abstract}

\maketitle

%
\section{Introduction}
%

Fundamentally, there are two ways to characterize a black hole.  The
first focuses on causal structure and says that a point in an
asymptotically flat spacetime is inside a black hole if no signal from
that point can reach future null infinity (see, for example,
\cite{hawkellis}). The boundary of the black hole region is the event
horizon. This is an intuitive definition but at the same time is
teleological and so highly non-local --- one must trace all causal
curves from a point before deciding whether or not it lies inside a
black hole.  By this definition neither large spacetime curvatures nor
singularities are necessary for black hole existence. 

In contrast, the second characterization is quasilocal and geometric,
saying that a point in spacetime is inside a black hole if it lies on a
trapped surface. Such surfaces, from which both future directed null
expansions are everywhere negative, are indicative of large spacetime
curvature and so this definition directly makes the association between
strong gravitational fields and black holes. It is also directly
connected to spacetime singularities as the existence of a trapped
surface is sufficient to imply the existence of a spacetime singularity
\cite{penrose}.\footnote{ Of course the two definitions are not entirely
independent. In spacetimes with a well-defined future null infinity,
trapped surfaces necessarily lie within the causally defined black hole
region. Further the two characterizations both identify the same region
for the family of Kerr-Newmann solutions \cite {hawkellis} (though this
is not necessarily true in more general spacetimes).}

Traditionally, the second viewpoint inspired the definition of an
\emph{apparent horizon}. Given a Cauchy surface, the trapped region 
is
defined as the (closure of the) union of all the trapped surfaces
contained in that slice of spacetime. The apparent horizon is then the
(two-dimensional) boundary of the trapped region \cite{hawkellis}.
Correspondingly, if a region of spacetime is foliated by Cauchy
surfaces, then one can locate the apparent horizon on each slice and 
so
define a time-evolved (three-dimensional) version of the apparent
horizon. Often this is also referred to as the apparent horizon. 

For any given foliation of a spacetime, most trapped surfaces will not
lie in the specified slices.  Thus the time-evolved apparent horizon is
defined by only a subset of the total number of trapped surfaces and 
so
is certainly slicing dependent and contained in the ``total" trapped
region. The time-evolved apparent horizons defined by various 
foliations
will typically intersect each other multiple times and also will usually 
have fully trapped surfaces lying partially outside of them (see for
example \cite {waldiyer, eardley,BadApVar} for discussions on these
points). 

By definition, an apparent horizon is a boundary between regions
containing trapped and untrapped surfaces.  As such, it is no surprise
that it is \emph{marginally outer trapped} --- that is the expansion,
$\theta_{(\ell)}$, of its outward null normal, $\ell$, vanishes
\cite{hawkellis}.  Now, while it is certainly not true that all such
surfaces will be apparent horizons, in practical terms it is clearly
easier to find the marginally outer trapped surfaces (MOTS) and then
identify apparent horizon candidates, rather than try to proceed by
first identifying all trapped surfaces. This is the approach taken in
numerical relativity (see for example \cite{thomas}) and in fact in that
field ``apparent horizon" is usually understood to mean the outermost
MOTS. 

Other properties can also be expected from an apparent horizon. First 
if
there are fully trapped surfaces ``just inside" the apparent horizon,
then there should exist arbitrarily small inward deformations generated
by some spacelike vector field $X^a$ under which the outward 
expansion
becomes negative, i.e. $\delta_X \tl < 0$.  Furthermore, by continuity
with the fully trapped surfaces, the expansion $\tn$ of the
inward-pointing null normal $n$ should be negative on the apparent
horizon.  Over a decade ago, Hayward formalized this intuition in his
definition of \emph{trapping horizons} \cite{hayward}. Here, we will be
mainly interested in future outer trapping horizons (FOTHs) which are
three-surfaces that may be foliated by two-surfaces with $\tl =0$, $\tn
< 0$ and $\delta_n \tl < 0$.  Although these will include most
time-evolved apparent horizons, the focus has now shifted to the
three-dimensional horizon surface itself.  There is no reference to a
foliation of the full spacetime. 

Marginally trapped surfaces are practical as they may be (relatively)
easily identified in simulations as well as studied with standard
geometric tools. However, they are also philosophically appealing as, in
contrast to event horizons, their evolution is causal.  As such, this
idea of studying the ``boundary" of a trapped region without first
finding a corresponding bulk has become increasingly popular in the last
few years. Apart from the further studies of trapping horizons
\cite{haywardPert, haywardFlux} there are closely related programmes
such as \emph{isolated horizons} which identify and study equilibrium
states \cite{isoPRL, LewIso, isoMech, isoGeom, BadIsoNum},
\emph{dynamical horizons}  which correspond to non-trivially evolving
horizons \cite{ak, theBeast, gregabhay, mttpaper, ams, GourgFlux,
Gourgoulhon:2006uc, KrishNum, bart, matt, MK}, and \emph{slowly evolving
horizons} which are ``almost isolated" trapping horizons \cite{prl,
billsPaper}. For reviews of the field see, for example \cite{akRev,
GourgRev, meRev}. 

Trapping and dynamical horizons each come equipped with a preferred
foliation into two-surfaces.  It is this foliation which is used to
define the null normals and hence the expansions.  In this paper we will
focus on variations of these two-surfaces as a route to a better
understanding of the existence, evolutions and deformations of the full
three-dimensional horizons. Given a vector field $X = A \ell - B n$
(where $A$ and $B$ are functions) normal to a horizon cross-section, the
corresponding variations are generated by $\delta_X$.  Particularly
important is the variation $\delta_X \tl$ which turns out to be a
second-order elliptic operator in $B$ that is defined by the intrinsic
and extrinsic geometry of the two-surface along with components of the
Einstein tensor.  The techniques used are similar to those described in
\cite{ams, Andersson:2005me}, however the emphasis is somewhat
different.  In those papers, the focus was on two-dimensional MOTS in a
three-dimensional slice of space-time, whereas we consider general
deformations of the two-surface in the full four-dimensional spacetime.

Solutions of $\delta_X \tl = 0$ generate both the evolution and the
possible deformations of FOTHs. A judicious application of a maximum
principle to the resulting elliptic partial differential equation is a
key to both deriving new results about these horizons and also 
unifying
much of the existing knowledge under a common formalism.  Among 
other
results, this technique will be used to show that any foliation of an
isolated FOTH may be freely deformed (a well-known result) while the
foliation of dynamical FOTH is rigid (a related version of this was
first shown in \cite{gregabhay}). Conversely we will see that an
isolated FOTH is rigid against normal deformations (it may only be
deformed into itself) while this is definitely not true for a dynamical
FOTH. However, the allowed deformations in the dynamical case are
strongly restricted by rules that are consistent with, although slightly
different from, those seen in \cite{gregabhay}.

Results from black hole physics also follow from the deformation
equations. Apart from the second law of horizon dynamics \cite {hayward,
ak, ams} and angular momentum flux laws \cite{by, ak, GourgFlux} we will
examine slowly evolving horizons in some detail \cite{prl, billsPaper}.
In particular we will explain how a horizon may be invariantly
characterized as ``almost" isolated. From this definition we will
examine the circumstances under which a FOTH has a well-defined and
slowly varying surface gravity and derive the first law for slowly
evolving horizons. The related flux laws for event and dynamical
horizons also follow from the variation equations. 

The plan of the paper is as follows.  We begin, in Section
\ref{twoSurf}, by considering the geometry of general spacelike
two-surfaces embedded in four-dimensional spacetimes and study 
how that
geometry changes if the surfaces are deformed . From there we 
specialize
to two-surfaces that satisfy $\tl = 0$, $\tn < 0$, and $\delta_n \tl<0$
and study the properties of deformations that preserve those 
conditions.
This is done in Section \ref{FOTS}. Next, in Section \ref{FOTHs}, we
apply these results to gain a better understanding of isolated,
dynamical, and future outer trapping horizons. With this general
understanding in hand we turn to a more specialized study of slowly
evolving horizons in Section \ref{SEH}. Finally we compare the flux 
laws
for slowly evolving horizons to the corresponding ones for dynamical 
and
event horizons in Section \ref{OtherFlux}.  Numerous technical results
are found in Appendices \ref{appA}--\ref{KerrFOTS}.

%
\section{Two-surfaces and their deformations}
\label{twoSurf}
%

To begin, we review the differential geometry of two-surfaces 
embedded
in four-dimensional spacetimes. Most of the results appearing in this
section are not new but it is useful to gather them together here both 
for
reference and to set the notation and emphasis that will be found in
future sections. 

\subsection{Two-surface geometry}
\label{twoSurfGeom}

Let $S$ be a closed and orientable two-surface that is (smoothly)
embedded in a four-dimensional time-oriented spacetime $(M,g_{ab})$
which has metric compatible covariant derivative $\nabla_a$.  Then 
there
are just two future-pointing null directions normal to $S$. Let $\ell^a$
and $n^a$ be null vector fields pointing in these directions; in
situations where this is meaningful we will always take $\ell^a$ and
$n^a$ as respectively outward and inward pointing.  If we further
require that $\ell \cdot n = -1$ then there is only one remaining degree
of rescaling freedom in the definition of these vector fields. 

The intrinsic geometry of $S$ is defined by the induced metric and
area-form. The definition of these quantities is independent of the
choice of null vectors above, however for our purposes it is most 
useful
to express them in terms of these vectors. Thus, the induced metric on
$S$ can be written as 
\bea
\tq_{ab} = g_{ab} + \ell_a n_b + \ell_b n_a \label{tq} \, , 
\eea 
while the area two-form $\vS$ satisfies $\boldsymbol{\epsilon} =
\boldsymbol{\ell} \wedge \boldsymbol{n} \wedge \vS$ where
$\boldsymbol{\epsilon}$ is the four-volume form on $M$.  This metric
also defines the compatible covariant derivative operator ${d}_a$ and
(two-dimensional) Ricci scalar $\tilde{R}$ on this two-surface. 

The extrinsic geometry describes how $S$ is embedded in $M$ and in 
the
usual way is defined by how the (in this case null) normal vectors
change over $S$. The extrinsic curvatures are:
\be
k^{(\ell)}_{ab} = \tq_a^c \tq_b^d \nabla_c \ell_d 
\quad  \mbox{and}  \quad
k^{(n)}_{ab} = \tq_a^c \tq_b^d \nabla_c n_d  \, . \label{kln}
\ee
These are symmetric since $\ell_a$ and $n_a$ are, by definition, 
surface
forming. In the standard way we decompose them as
\bea
k^{(\ell)}_{ab} = \frac{1}{2} \tl \tq_{ab} + \sigma^{(\ell)}_{ab} 
\quad \mbox{and} \quad 
k^{(n)}_ {ab} = \frac{1}{2} \tn \tq_{ab} + \sigma^{(n)}_{ab} \, . \label{dq}
\eea 
where the \emph{expansions}
\begin{eqnarray} \theta_{(\ell)} = \tq^{ab} \nabla_{a} \ell_{b} 
  \quad \mbox{ and } \quad 
  \theta_{(n)} = \tq^{ab} \nabla_{a} n_{b} \, , \end{eqnarray} 
are the traces of the extrinsic curvatures and the \emph{shears} 
\begin{equation}\label{shears}
   \sigma^{(\ell)}_{ab} \equiv \left(\tilde{q}_a^c \tilde{q}_b^d 
   - \frac{1}{2} \tilde{q}_{ab}\tilde{q}^{cd} \right) \nabla_{c} \ell_{d}  
   \quad \mbox{and} \quad 
   \sigma^{(n)}_{ab} \equiv \left(\tilde{q}_a^c
   \tilde{q}_b^d - \frac{1}{2} \tilde{q}_{ab}\tilde{q}^{cd}
   \right) \nabla_{c} n_{d}  \, ,
\end{equation}
are the trace-free parts.

The last part of the extrinsic geometry is given by the connection on
the normal cotangent bundle $T^*_{\mspace{-6mu} \perp} \mspace
{-2mu} S$
namely
\begin{equation}\label{omega}
  \tilde{\omega}_{a} := - \tq_a^b n_{c} {\nabla}_{b} \ell^{c} \, .
\end{equation}
To see that this is the connection consider a general normal one-form
$\mu_a = \alpha \ell_a + \beta n_a \in T_{\mspace{-6mu} {\perp}}^* 
\mspace{-2mu} S$.
Then a direct calculation shows that
\bea
\tq_a^c \nabla_c \mu_b = k^{(\mu)}_{ab} 
+ ({d}_a \alpha + \tom_a \alpha) \ell_b 
+ ({d}_a \beta - \tom_a \beta) n_b \, \, , \label{dX}
\eea
where $k^{(\mu)}_{ab} = \alpha k^{(\ell)}_{ab} + \beta k^{(n)}_{ab}$.
Thus, the covariant derivative on this normal bundle is
\bea
d^{^\perp}_a  (\alpha \ell_b + \beta n_b ) :=  
({d}_a \alpha + \tom_a \alpha) \ell_b + ({d}_a \beta - \tom_a \beta) n_b 
\, , 
\eea
and $\tom_a$ clearly acts as the connection.  The gauge dependence 
in
this case is the scaling chosen for the null vectors.  If $\ell
\rightarrow f \ell$ and $n \rightarrow n/f$ for some function $f$, then
the corresponding transformation for the connection is  
\bea
\tom_a \rightarrow \tom_a + d_a \ln f \, \, .\label{tom}
\eea
As usual the geometric, gauge independent, information associated 
with
the connection is contained in its curvature which in this case is
\bea
\Omega_{ab} = d_a \tom_b - d_b \tom_a \label{Om} \, , 
\eea
and this is constrained by the four-space curvature via the Ricci
equation (Appendix \ref{Prel}):
\bea
\Omega_{ab}
= \tq_a^c \tq_b^d \ell^e n^f C_{cdef} 
+ \sigma^{(\ell) c}_a \sigma^{(n)}_{bc} 
- \sigma^{(\ell) c}_{b} \sigma^{(n)}_{ac} \, , 
\eea
where $C_{cdef}$ is the Weyl curvature of the full spacetime. In this
paper we will usually be more interested in the connection itself rather
than this curvature. 

Other constraints relating the geometry of $S$ to the full four-space
curvature come from the Gauss and Codazzi equations. The Gauss 
equation is  
\bea
\tq_a^e \tq_b^f \tq_c^g \tq_d^h \mathcal{R}_{efgh} 
= \tilde{R}_{abcd}
+ (k^{(\ell)}_{ ac}  k^{(n)}_{ bd} + k^{(n)}_{ ac}  k^{(\ell)}_{ bd}) 
-  (k^{(\ell)}_{ bc}  k^{(n)}_{ ad} + k^{(n)}_{ bc}  k^{(\ell)}_{ ad}) \, \, , 
\eea
where $\mathcal{R}_{efgh}$ and $\tilde{R}_{abcd}$ are the (four- and
two-dimensional) Riemann tensors, while the (slightly modified) 
Codazzi
equations are
\begin{eqnarray}
  ({d}_{a} - \tom_a) \theta_{(\ell)} &=&
  2 ({d}_{b} - \tom_b) \sigma^{(\ell) b}_{\; a}
  - \tilde{q}_{a}^{b}  G_{bc}\ell^{c} 
    - 2 \tq_a^b  C_{bcde} \ell^{c} \ell^{d} n^{e} \label{expl2deriv} 
    \mbox{ and} \\
  ({d}_{a} + \tom_a) \theta_{(n)} &=& 
  2 ({d}_{b} + \tom_b) \sigma^{(n) b}_{a} 
  - \tilde{q}_{a}^{b} G_{bc} n^{c} 
    + 2 \tq_a^b C_{bcde} n^{c} \ell^{d} n^{e}  \, , \label{expn2deriv}
\end{eqnarray}
where $G_{ab} = \mathcal{R}_{ab} - \frac{1}{2} \mathcal{R} g_{ab}$ is
the Einstein tensor.  A derivation of these relations can be found in
Appendix \ref{appA}.

\subsection{Deforming a two-surface}
\label{twoSurfDef}

\subsubsection{Defining variations}

\begin{figure}
\includegraphics{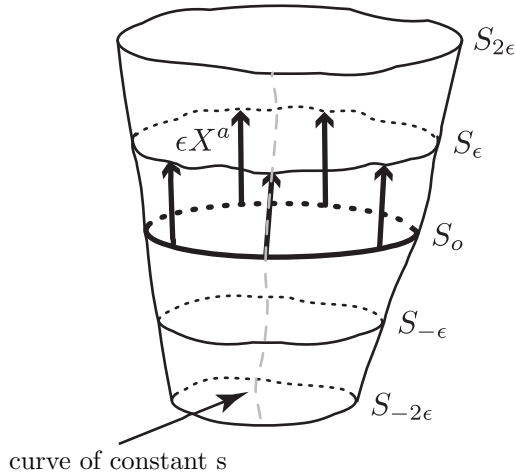}
\caption{A schematic of a section of $\mathcal{T}_\Phi$ around $S_o$.
The variation vector field $X^a$ is everywhere tangent to the tube,
points along curves of constant $s$,  and generates the foliation. Thus
for sufficiently small $\epsilon$, one can intuitively write $S_\epsilon
= S_o + \epsilon X$.}
\label{VarSurf}
\end{figure}

\begin{figure}
\includegraphics{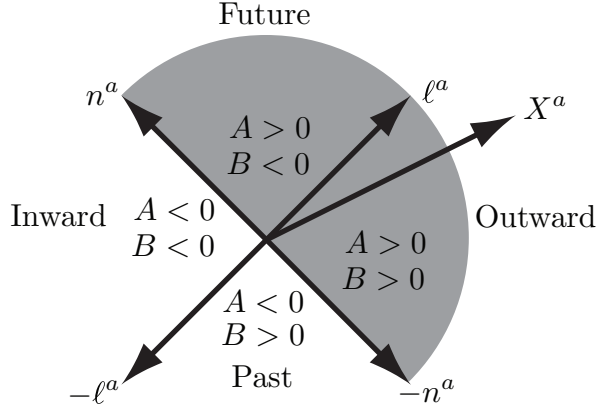}
\caption{The normal space at a point $s \in S_\lambda$. Normal 
vectors
written as $X^a = A \ell^a - B n^a$ point into the causal future if
$A\geq 0$ and $B \leq 0$. The shaded region represents the
$\ell$-oriented vector fields for which $A>0$ and $B$ may take any
value.}
\label{FO}
\end{figure}

A \emph{variation} or \emph{deformation} of a two-surface $S_o$ is a
smooth, one-to-one function $\Phi (s,\lambda): S_o \times [-
\lambda_o,
\lambda_o] \rightarrow M$ (with $\lambda_o$ some real number) such 
that
$\Phi(S_o,0) = S_o$. Thus, $\Phi$ generates a (finite) three-surface
$\mathcal{T}_\Phi$ and that surface is foliated by images $S_\lambda 
=
\Phi(S_o,\lambda)$ of $S_o$ as depicted in Fig.\ \ref{VarSurf}. The
\emph{variation vector field} $X^a = (\partial/\partial \lambda)^a$ is
tangent to the curves of constant $s\in S_o$. The flow generated by 
this
vector field maps leaves of constant $\lambda$ into each other. Unless
otherwise noted, we will restrict our attention to \emph{normal
variations} where $X^a$ is everywhere perpendicular to the $S_ 
\lambda$
and so can be written: 
\bea
X^a = A \ell^a - B n^a \, , \label{Xdef}
\eea
for some functions $A$ and $B$. There are no restrictions on the values
of $A$ and $B$. However, in later sections we will usually assume that
$X^a$ is \emph{$\ell$-oriented} so that $A>0$.  Then if $B>0$, $X^a$ is
spacelike while $B<0$ means that it is timelike. We will mainly be
interested in situations where $B>0$, hence the negative sign in
(\ref{Xdef}). Fig.\ \ref{FO} should help to keep the various cases
straight. 

For any values of $A$ and $B$, the map $\Phi$ deforms $S_o$ into
successive surfaces $S_\lambda$.  To quantify the change in the
geometry, we identify points along the curves of constant $s$ and then
calculate derivatives of the intrinsic and extrinsic geometry with
respect to the parameter $\lambda$. Taking the normal connection
$\tom_a$ as an example, we write its variation as $\delta_X \tom_a$ 
and
note that calculating this quantity amounts to:
\begin{enumerate}

\item using $\Phi$ to construct the $S_\lambda$ in a neighbourhood of
$\lambda = 0$, 

\item constructing the $\tom_a$ on those $S_\lambda$ (among other 
things
this will involve choosing a scaling of the null vectors), and

\item calculating the Lie derivative $\Lie_X \tom_a$ and pulling-back
the result onto $S_o$.

\end{enumerate}
Thus, once the two-surfaces are constructed and $\tom_a$ calculated, 
we have 
\bea
\delta_X \tom_a = \tq_a^c \Lie_X \tom_c \, , 
\eea
in standard four-dimensional notation. 

\subsubsection{Calculating variations}

We now calculate some of these variations for the geometric quantities
that are of interest in this paper. The easiest calculation is the
variation of the two-metric. A couple of lines of algebra shows 
\bea
\delta_X \tq_{ab} = A k^{(\ell)}_{ ab} - B k^{(n)}_{ ab} 
=  (A \tl - B \tn) \tq_{ab} 
+  2 (A \sigma^{(\ell)}_{ ab} - B \sigma^{(n)}_{ ab})  \label{Lq} \, ,
\eea
from which it follows that 
\bea
\delta_{X} \vS = (A \tl - B \tn) \vS \, . \label{LcV} 
\eea
These expressions justify referring to $\tl$ and $\tn$ as expansions 
and
$\sigma^{(\ell)}_{ab}$ and $\sigma^{(n)}_{ab}$ as shears; respectively
they describe how $S_o$ expands and shears if deformed in the null
directions. 

Finding the variations of the extrinsic quantities is more involved.
Here we just outline the calculations but more details can be found in
Appendix \ref{DefEq}. First we note that since $\ell_a$ and $n_a$ are
everywhere normal to the $S_\lambda$, both $\delta_X \ell_a = 0$ and
$\delta_X n_a = 0$ (with the usual pull-backs understood). Thus
\bea
X^b \nabla_b \ell_a & = &  - ({d}_a  - \tom_a) B + \kappa_X \ell_a
\,  \mbox{ and} \label{LXl} \\
X^b \nabla_b n_a & = & ({d}_a  +  \tom_a)A  - \kappa_X n_a \, \,  \label
{LXn}  
\eea
where
\bea 
\kappa_X = - X^a n_b \nabla_a \ell^b \,  \label{surfGrav}
\eea
is the component of the connection on the $S_\lambda$ normal
bundles in the $X$-direction.  Explicitly, under rescalings $\ell
\rightarrow f \ell$ and $n \rightarrow n/f$,
\bea
\kappa_X \rightarrow \kappa_X + \Lie_X \ln f  \label{kappaX} \, \, . 
\eea
We will usually refer to $\kappa_X$ as the \emph{surface gravity}
associated with $X^a$ in analogy with the corresponding quantities 
on a Killing or isolated horizon (though at this stage we make no
claims about the physical content of this nomenclature). 

The importance of (\ref{LXl}) and (\ref{LXn}) is that they allow us to
convert expressions involving derivatives off of $S_o$ into ones 
containing only quantities
defined on $S_o$ plus the gauge dependent surface gravity $
\kappa_X$.
Then, with the help of these relations a direct calculation shows that 
\begin{eqnarray} 
   \delta_X \theta_{(\ell)} - \kappa_X \tl &=& 
   - {d}^{\, 2} \mspace{-2mu} B  
   + 2 \tilde{\omega}^{a} {d}_{a} B 
   - B \left[ \norm \tilde{\omega} \norm^2
     -  {d}_{a} \tilde{\omega}^{a} 
     - \tilde{R} /2
     + G_{ab} \ell^{a} n^{b} - \tl \tn \right] \nn \\
     & &    \quad    - A \left[ \norm\sigma^{(\ell)}\norm^{2} 
     + G_{ab} \ell^{a} \ell^{b}
     + (1/2) \theta_{(\ell)}^{2} \right] \label{explderiv} \, \, , 
\end{eqnarray}
where $\norm \tom \norm^2 = \tom_a \tom^a$ and  $\norm \sigma^
{(\ell)}
\norm^2 = \sigma^{(\ell)}_{ ab} \sigma^{(\l) ab}$.  

We will also need to know $\delta_X \tn$ and this is most easily
calculated by substitutions into the above expression (\ref{explderiv}).
Exchanging $\ell_a$ and $n_a$ and sending $A \rightarrow -B$ and 
$B
\rightarrow - A$  it is straightforward to see that 
\bea
  \delta_X \theta_{(n)} + \kappa_X \tn &=& 
  {d}^{\, 2} \mspace{-2mu} A 
    + 2 \tilde{\omega}^{a} {d}_{a} A 
    + A \left[\norm\tilde{\omega}\norm^2  
    + {d}_{a} \tilde{\omega}^{a} - \tilde{R}/2 
    + G_{ab} n^{a} \ell^{b} - \tl \tn \right] \nn \\
  && \quad 
  + B \left[ \norm\sigma^{(n)}\norm^{2} 
    + G_{ab} n^{a} n^{b} 
    + (1/2)\theta_{(n)}^{2} \right] \,   \label{expnderiv}
      \, .
\eea
Finally, the variation of the normal connection (also derived in
Appendix \ref{DefEq}) is
\begin{eqnarray}
\label{omegaderiv}
  \delta_X \tilde{\omega}_{a}  - {d}_a \kappa_X  &=& 
    -  k^{(\ell)}_{ a b} \left[ {d}^b A  +  \tilde{\omega}^b A \right]
    - k^{(n)}_{ ab} \left[{d}^b B - \tilde{\omega}^b B 
    \right] \\
  & &   + \tilde{q}_{a}^{\; \; b} \left[ \frac{1}{2} G_{bc}\tau^{c}
    - C_{bcde} {X}^{c}\ell^{d}n^{e} \right]    \nonumber \, , 
\end{eqnarray}
where $\tau^c = A \ell^c + B n^c$ is normal to $X^c$. Enlisting the help
of the Codazzi equations  by combining $A \times (\ref{expl2deriv}) + B
\times (\ref{expn2deriv}) - 2 \times (\ref{omegaderiv})$ this can be
rewritten as 
\begin{eqnarray}\label{ang_mom_expression1}
  \delta_{X} \tilde{\omega}_{a} + ( A \tl - B \tn) \tom_a 
    &=&  {d}_{a} \kappa_{X} 
    -  {d}_{b} \left( A \sigma_{(\ell) \; a}^{\; \; b}  
    + B \sigma_{(n) \; a}^{\; \; b} \right) 
    + \tilde{q}_{a}^{\; b} G_{bc} \tau^{c} \\
    && \quad  + \frac{1}{2} {d}_{a} \left( A \theta_{(\ell)} 
    + B \theta_{(n)} \right) - \theta_{(\ell)} {d}_{a} A
    - \theta_{(n)} {d}_{a} B  \nonumber  \, , 
\end{eqnarray}
which eliminates the Weyl dependence. This is the form that we will 
use. 

These variations will be sufficient for most of our considerations. Note
that equivalent or closely related versions of the expressions for
$\delta_X \tl$, $\delta_X \tn$, and $\delta_X \tom_a$ have previously
appeared in, for example, \cite{GourgFlux, MK, ams, eardley, meRev,
HaywardDualNull}, though not with the particular two-surface emphasis
that we adopt here. 

\subsection{Angular momentum and its evolution}

\label{angMom}

Physically, the connection $\tom_a$ defines the angular momentum
associated with any \emph{rotation vector field} $\phi^a$ on a closed
two-surface $S$ \cite{by, ak, prl, GourgFlux, KrishNum, haywardFlux}.
By definition the flow associated with such a $\phi^a$ has only two
fixed points and foliates the remainder of $S$ into closed integral
curves of parameter length $2 \pi$. The canonical example of such a
field is a Killing vector field of the two-metric $\tq_{ab}$ and in that
particular case it replaces the flat-space notion of an axis of rotation.
However even if it is not a Killing vector field, it is standard to
define the angular momentum of $S$ relative to $\phi^a$ as 
\bea
\label{am1}
J[\phi] = \frac{1}{8 \pi G} \int_{S} \vS \phi^a \tom_a \, \, . 
\eea
Note that any rotation vector field $\phi^a$ is necessarily divergence
free and a quick calculation with the help of (\ref{tom}) shows that
this expression is independent of the scaling of the null vector fields.
Alternatively if the surface $S$ has a suitable topology such as $S^2$,
then a divergence free $\phi^a$ can necessarily be written in the form
$\phi^a = \tilde{\epsilon}^{ab} d_b \zeta$ 
for the area-form $\tilde{\epsilon}_{ab}$ and
some function $\zeta$.  Then, the scaling independence is made 
explicit
if we rewrite (\ref{am1}) as  
\bea
J[\phi] = \frac{1}{8 \pi G} \int_{S} \mspace{-6mu} \zeta  
  \mathbf{\Omega} \, , 
\eea
where as usual we drop the indices and write in bold any form that is being
integrated over. 

We now consider how the angular momentum changes under
deformations. To this end we multiply (\ref{ang_mom_expression1}) by 
the
area form $\vS$ and so obtain 
\begin{eqnarray}\label{ang_mom_expression}
  \delta_{X} \left( \vS \tilde{\omega}_{a} \right) 
     &=& \vS \left( {d}_{a} \kappa_{X} 
      - {d}_{b} \left( A \sigma^{(\ell) b}_{a}  
      + B \sigma^{(n) b}_{\; a} \right) 
      + \tilde{q}_{a}^{\; b} G_{bc} \tau^{c} \right)  \\
    &&\qquad  + \vS \left(  \frac{1}{2} {d}_{a} \left( A \theta_{(\ell)} 
      + B \theta_{(n)} \right) 
      - \theta_{(\ell)} {d}_{a} A
      - \theta_{(n)} {d}_{a} B \right) \, . \nonumber
\end{eqnarray}
Now, extend $\phi^a$ off of $S$ by demanding that $\delta_X \phi^a = 
0$
--- essentially this is equivalent to the flat space requirement that
angular momentum be measured relative to a fixed axis of rotation. 
Then,
contracting (\ref{ang_mom_expression}) with $\phi^a$ and integrating
over $S$, several total divergence terms vanish and we find that 

\begin{equation}\label{ang_mom_evolution}
  \delta_X J[\phi]  = \frac{1}{8 \pi G} \int_{S} \vS \left\{  
    G_{ab} \phi^{a} \tau^{b} 
    + \sigma^{({\tau})} \mspace{-6mu} : \mspace{-4mu} \sigma^{(\phi)} 
+ A \phi^a d_a \tl + B \phi^a d_a \tn \right\} \, , 
\end{equation}
where the shear with respect to a general vector field $w^a$ is 
\begin{equation}\label{general_shear}
   \sigma_{ab}^{(w)} \equiv 
     \left(\tilde{q}_a^c \tilde{q}_b^d 
     - \frac{1}{2} \tilde{q}_{ab}\tilde{q}^{cd} \right) 
     \nabla_{c} w_{d}  \, , 
\end{equation}
and $\sigma^{({\tau})}  \mspace{-6mu} : \mspace{-4mu} \sigma^{(\phi)} 
=
\sigma^{(\tau)}_{ab} \sigma^{(\phi) ab}$ (this double contraction
notation is adopted from \cite{GourgFlux}). 

We will return to equation (\ref{ang_mom_evolution}) in
Section\ref{SEH_angMom} where we will be able to neglect the last
two-terms and so interpret the change in angular momentum as coming from
a flux of stress-energy (with the help of the Einstein equation) and a
flux of shear. A detailed discussion and fluid mechanical interpretation
of (\ref{ang_mom_expression}) and (\ref{ang_mom_evolution}) can also be
found in \cite{GourgFlux}. 

\subsection{The ``constraint" law}

Finally, before specializing to the two-surfaces associated with 
horizons we 
derive the following relation. First, combining 
$A \times (\ref{explderiv}) + B \times (\ref{expnderiv})$ 
we find that 
\bea
\kappa_{X} \theta_{X} &=& A \delta_X \tl + B \delta_X \tn 
      + d_a (A d^a B - B d^a A - 2AB \tom^a) \label{kapT} \\
           & & 
   + \sigma^{(\tau)} \mspace{-6mu} : \mspace{-4mu} \sigma^{(X)} 
   + G_{ab} X^a \tau^b 
   +  \frac{1}{2} \theta_{(X)} \theta_{(\tau)} \, , \nn
\eea
where we again have $\tau_a = A \ell_a + B n_a$ and $\theta_{(X)}$ 
and
$\theta_{(\tau)}$ are defined in the obvious way.  Integrating this over
$S$, the total divergence term vanishes and we find that
\begin{eqnarray}\label{constraint_non_rot}
  \frac{1}{8 \pi G} \int_{S} \kappa_{X} \delta_X \tilde{\epsilon}
  &=& \frac{1}{8 \pi G} \int_{S} \tilde{\epsilon} \left[ 
    G_{ab} X^a \tau^b 
    + \sigma^{(X)}  \mspace{-6mu} : \mspace{-4mu} 
    \sigma^{(\mathcal{\tau})} \right] \nonumber \\
  && \qquad +\frac{1}{8 \pi G} \int_{S} \tilde{\epsilon} \left[ 
   A \delta_X \theta_{(\ell)} +
    B \delta_X \theta_{(n)}
    + \frac{1}{2} \theta_{(X)} \theta_{(\mathcal{\tau})} \right] \, . 
\end{eqnarray}
More generally for a vector field 
\bea
\mathcal{X}^a = X^a + \tilde{x}^a \, , 
\eea
where $\tilde{x}^a$ is everywhere transverse to the $S$ we can
combine (\ref{ang_mom_expression}) and (\ref{kapT}) to obtain 
\begin{eqnarray} \label{pre_constraint}
  \frac{1}{8 \pi G} \int_{S} \left\{ \kappa_{X} \delta_X \vS +
    \tilde{x}^{a} \delta_X 
      \left( \tilde{\epsilon} \tilde{\omega}_{a} \right) \right\}
  &=& \frac{1}{8 \pi G} \int_{S} \tilde{\epsilon} \left[ 
    G_{ab} \mathcal{X}^a  \tau^{b} 
    + \sigma^{(\mathcal{X})}  
    \mspace{-6mu} : \mspace{-4mu} \sigma^{(\mathcal{\tau})} 
    \right] \\
  && \qquad +\frac{1}{8 \pi G} \int_{S} \tilde{\epsilon} \left[ 
   A \delta_{\mathcal{X}} \theta_{(\ell)} +
    B \delta_{\mathcal{X}} \theta_{(n)}
    + \frac{1}{2} \theta_{(X)} \theta_{(\mathcal{\tau})} \right] \, . \nn
\end{eqnarray}

In cases where $\kappa_X$ is constant and $\tilde{x}^a = \Omega 
\phi^a$
for some constant $\Omega$ (which is not related to the curvature of 
the
normal bundle) and rotation vector field satisfying $\delta_X \phi^a =
0$, the left-hand side of this equation takes a particularly familiar
form: \bea \frac{\kappa_X}{8 \pi G} \dot{a} + \Omega \dot{J}[\phi] \, \,
.  \eea The similarity to the first law of black hole mechanics is not
coincidence. In sections \ref{SEH} and \ref{OtherFlux}  we will see that
the dynamical version of the first law for both event and trapping
horizon are closely related to (\ref{pre_constraint}).

A version of this relation was referred to as the horizon constraint law
in \cite{prl, theBeast} due to its equivalence to the integrated
diffeomorphism constraint on $\mathcal{T}_\Phi$. A discussion of its
terms and their interpretation if the horizon is viewed as a viscous
fluid can also be found in \cite{GourgFlux}.  Furthermore, in
\cite{Gourgoulhon:2006uc}, equation (\ref{pre_constraint}) is
interpreted as a second order evolution equation for the area.

\section{Future outer trapped surfaces}
\label{FOTS}

The expressions of the previous section hold for any spacelike
two-surface embedded in any four-dimensional spacetime. In this 
paper
however, our main interest will in the two-surfaces that foliate future
outer trapping horizons which in turn are embedded in solutions of the
Einstein equations. Thus in this section we consider spacelike
two-surfaces on which $G_{ab} = 8 \pi T_{ab}$ and for which $\tl = 0$,
$\tn < 0$ and there is a scaling of the null vectors such that $\delta_n
\tl < 0$.  Adapting Hayward's nomenclature we will call such surfaces 
\emph{future outer trapping surfaces (FOTS)}. 

First, we consider the conditions under which a \emph{marginally 
trapped
surface} $S$ with $\tl = 0$ and $\tn < 0$ is a FOTS. To this end, we we
set $A=0$, $B=-1$ and $\tl=0$ in (\ref{explderiv}) and so find that on a
marginally trapped surface
\bea
\delta_n \tl  = - \tilde{R} /2 +  \norm \tilde{\omega} \norm^2 
     -  {d}_{a} \tilde{\omega}^{a} 
     + (8 \pi G) T_{ab} \ell^{a} n^{b}  \, .  \label{dntl}
\eea
Then it is immediate that the $\delta_n \tl < 0$ condition is determined
entirely by this component of the stress-energy tensor  along with the
intrinsic and extrinsic geometry of $S$ -- no derivatives need to be
taken off of the surface. That said, checking this condition is slightly
more complicated than just calculating this quantity with an arbitrary
scaling of the null vectors. For example, in Appendix \ref{KerrFOTS} it
is shown that for the standard scaling of null vectors on a Kerr
horizon, $\delta_n \tl$ is not always less then zero. A rescaling is
necessary for this relationship to become apparent. 

Now, for any spacelike two-surface on which $\tl = 0$, equation
(\ref{explderiv}) simplifies to become 
\begin{eqnarray}
 \delta_{X} \theta_{(\ell)} = 
   - {d}^{\, 2} \mspace{-2mu} B  
   + 2 \tilde{\omega}^{a} {d}_{a} B 
   - B \delta_n \tl  
   + A  \delta_\ell \tl \, , \label{Lvt}
\end{eqnarray}
where $\delta_n \tl$ takes the form shown in equation (\ref{dntl}) and
\begin{eqnarray}
\delta_\ell \tl &=& - \norm \sigma^{(\ell)} \norm^{2} 
     - (8 \pi G) T_{ab} \ell^{a} \ell^{b} \,  \, .    \label{dltl}
\end{eqnarray}
If this is a FOTS we can adopt a scaling so that $\delta_n \tl <
0$, while if we assume the null energy condition it also follows that
$\delta_\l \tl \leq 0$. Then, we can draw several conclusions about
FOTS. 

First strengthening to the dominant energy condition, there is the
well-known \cite{newman} restriction on the topology of such a 
surface. 

\newtheorem{prop}{FOTS Property}

\begin{prop}

If $S$ is a closed and orientable FOTS on which the dominant energy
condition holds, then it is homeomorphic to $S^2$. \label{FOTStop}
\end{prop} 

\noindent This follows from equation (\ref{dntl}). On integrating this
over $S$ and doing a bit of rearranging we find that : 
\bea
\chi = \frac{1}{2\pi} \int_{S} \tilde{\epsilon} \left\{- \delta_n \tl 
  + \norm\tilde{\omega}\norm^2 + 8 \pi T_{ab} \ell^a n^b  \right\} \, ,
\label{top}
\eea
where $\chi$ is the Euler characteristic of $S$. Now, by assumption
$\delta_n \tl < 0$ while the dominant energy condition implies that the
matter term is positive. Thus $\chi>0$ and this is sufficient to tell us
that $S$ must be homeomorphic to a two-sphere since that is the only
closed and orientable two-surface with positive Euler characteristic.
$\Box$

Next, we consider the conditions under which a FOTS may be 
deformed
whilst preserving its defining characteristics. Sufficiently small
variations will always leave $\tn < 0$ and $\delta_n \tl < 0$ and so the
key to understanding these deformations is finding normal vector fields
$X^a$ such that $\delta_X \tl = 0$. We assume that all fields are at
least twice differentiable. 

We begin with the case where $\delta_\ell \tl = 0$ everywhere on a 
FOTS.
Then we have: 

\begin{prop}\label{Adef0}

If $S$ is a FOTS on which $\delta_\ell \tl = 0$ everywhere, then
variation vectors $X^a$ satisfy $\delta_X \tl = 0$ if and only if they
are parallel to $\ell^a$. 
\end{prop}

\noindent Starting with the first part, if $\delta_\ell \tl = 0$
everywhere on $S$ then equation (\ref{Lvt}) becomes
\begin{eqnarray} 
\delta_X \tl  =  {d}^{\, 2} \mspace{-2mu} B  
   - 2 \tilde{\omega}^{a} {d}_{a} B 
   + B \delta_n \tl  \, \, ,  \label{isoEq}
\end{eqnarray}
and it is trivial that $B = 0 \Rightarrow \delta_X \tl = 0$ for any
value of $A$.  It is also straightforward to see that the converse must
be true. First applying the maximum principle of Appendix \ref{maxmin}
to $\delta_X \tl = 0$ we find that $B$ must be either constant or
everywhere negative. Similarly applying the corresponding minimum
principle to $-\delta_X \tl = 0$ we find that $B$ must be either
constant or everywhere positive. Thus, $B$ must be constant and it is
clear that since $\delta_n \tl \neq 0$ that constant must be zero. The
result is established.
$\Box$

It is also true that all such deformations leave the intrinsic geometry of 
$S$ invariant:

\begin{prop}\label{FOTSgeom}

Let $S$ be a FOTS on which $\delta_\ell \tl = 0$ everywhere and the 
null
energy condition holds. Then $\delta_X \tl = 0$ implies that $\delta_X
\tq_{ab} = 0$. That is, all deformations leave the intrinsic geometry
invariant. 

\end{prop}

\noindent If the null energy condition holds then all terms in
(\ref{dltl}) are non-negative and so if $\delta_\ell \tl = 0$ they must
all, including $\sigma^{(\ell)}_{ab}$, be zero. Then with $B = 0$ by
Property \ref{Adef0} and $\tl = 0$ by assumption, equation (\ref{Lq})
implies that $\delta_X \tq_ {ab} = 0$ as required.  $\Box$

Such results are familiar from the isolated horizon literature and we
will return to them in section \ref{FOTHs}. For now however we 
consider
FOTS on which $\delta_\ell \tl$ is somewhere non-zero. In doing this 
we
restrict our attention to $\ell$-oriented variation vector fields for
which $A> 0$ (Fig.\ \ref{FO}).

\begin{prop}\label{dynamic}
Let $S$ be a FOTS and assume the null energy condition.  Then, if
$\delta_\ell \tl \neq 0$ anywhere on $S$, all $\ell$-oriented variation
vectors $X^a$ that satisfy $\delta_X \tl = 0$ are spacelike everywhere
on $S$. 
\end{prop}

\noindent If the null energy condition holds then $\delta_\ell \tl \leq
0$ by equation (\ref{dltl}).  Thus with $A > 0$,  equation (\ref{Lvt})
implies that 
\bea
- {d}^{\, 2} \mspace{-2mu} B  
   + 2 \tilde{\omega}^{a} {d}_{a} B 
   - B \delta_n \tl  \geq 0 \, ,  \label{Op}
\eea
everywhere on $S$. Then by the minimum principle of Appendix
\ref{maxmin},  $B$ is either everywhere positive or everywhere 
constant.
If it is constant then the derivatives in (\ref{Lvt}) vanish and 
\bea
B  = A \left( \frac{\delta_\ell \tl}{\delta_n \tl} \right)  \, . 
\eea
everywhere on $S$. In particular this must hold at the point where
$\delta_\ell \tl < 0$ and so with $A>0$ and $\delta_n \tl < 0$ we again
find that $B > 0$. Thus in either case $g_{ab} X^a X^b = 2AB > 0$ and
$X^a$ is spacelike.  $\Box$

Combining this with Property \ref{Adef0} we see that FOTS satisfying 
the
null energy condition may be cleanly split into two classes -- those for
which $\ell$-oriented variation vector fields satisfying $\delta_X \tl =
0$ are null and those for which these vectors are spacelike. Such an
$X^a$ cannot be timelike and what is more it cannot be partly null and
partly spacelike. 

We also know something about how the geometry of a FOTS must 
change with
respect to a spacelike $\tl = 0$ preserving deformation:

\begin{prop}\label{Dyngeom}
Let $S$ be a FOTS and assume the null energy condition. Then if
$\delta_\ell \tl$ is non-zero anywhere on $S$, $\delta_X \vS > 0$
everywhere. The deformation causes $S$ to expand everywhere.  
\end{prop} 

\noindent If $\delta_\l \tl \neq 0$ anywhere on $S$, then by Property
\ref{dynamic}, $X^a$ must be everywhere spacelike with $B>0$. Then
$\delta_X  \vS = - B \tn \vS$ where $- B \tn > 0$ since $\tn < 0$ by
assumption.  $\Box$

Finally, it is quite clear that in general there will be an infinite
number of $X^a$ that will solve $\delta_X \tl = 0$ and so an equally
infinite number of FOTS-preserving deformations. For example, if
$\delta_\ell \tl$ is nowhere zero and we choose \emph{any} $B \in
C^2(S)$ then we can always solve $\delta_X \tl = 0$ to find a
corresponding $A$ (though unless $B$ is constant, there is no 
guarantee
that the resulting $X^a$ will be $\ell$-oriented). Furthermore if
$\delta_\l \tl = 0$ everywhere, then Property \ref{Adef0} tells us that
for \emph{any} $X^a = A \ell^a$, $\delta_X \tl = 0$.

\begin{figure}
\includegraphics{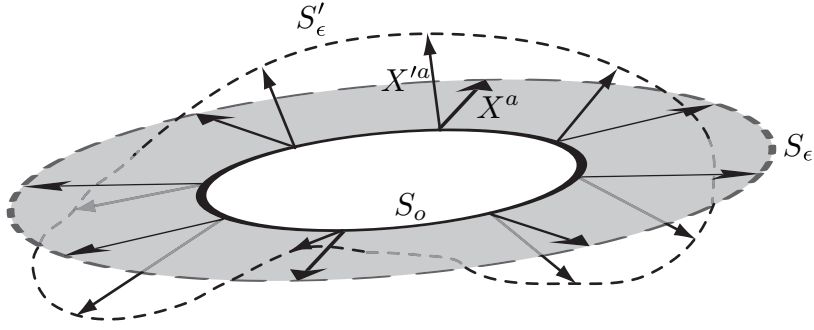}
\caption{Given two spacelike variation vector fields $X^a$ and $X'^a$ 
on
a FOTS that satisfy $\delta_X \tl = \delta_{X'} \tl = 0$, then if they
are not constant multiples of each other they must interweave so that
$X'^a$ will sometimes point into the future of $X^a$ and sometimes 
will
point into the past.}
\label{TwoVars}
\end{figure}

In contrast to the $\delta_\ell \tl = 0$ case where all allowed
variation vector fields must be parallel to $\ell^a$, if $\delta_\l \tl$
is somewhere non-zero then (apart from constant rescalings) no two
variation vector fields are parallel. Instead they must interweave as
shown in Fig.\ \ref{TwoVars}. 

\begin{prop} \label{TwoVarsProp}
Let $S$ be a FOTS with $\delta_\l \tl$ somewhere non-zero and 
assume the
null energy condition. Further let $X^a$ and $X'^a$ be two
$\ell$-oriented, FOTS-preserving deformation vector fields. Then either

\begin{enumerate}
\item $X'^a = \lambda_o X^a$ for some constant $\lambda_o$ or 

\item $X'^a$ interweaves $X^a$ in the sense that $X'^a \tau_a$ takes
both positive and negative values on $S$, where $\tau^a = A \ell^a + B
n^a$ is the usual forward pointing timelike normal to $S$. 

\end{enumerate}
\end{prop}

\noindent By Property \ref{dynamic}, both $X^a$ and $X'^a$ are 
spacelike
and so with $X^a = A \ell^a - B n^a$ for some positive $A$ and $B$, 
we
have
\bea
X'^a  = (\alpha + \lambda) A \ell^a - \lambda B n^a
\eea
for some functions $\alpha$ and $\lambda$ that satisfy both $\lambda 
>
0$ and $\alpha + \lambda >0$. 

Now, to begin our analysis, let us consider the case where $\alpha$
takes both positive and negative values. Then we immediately see that
\bea
X'^a \tau_a = - \alpha AB \, , 
\eea
and so this corresponds to the second posited behaviour for $X'^a$. 

By contrast, if $\alpha$ does not take both positive and negative 
values
then at least one of $\alpha \leq 0$ or $\alpha \geq 0$ must be true. As
preparation to exploring these two possibilities, we note that given
$\delta_X \tl = 0$, it is straightforward to see that $\delta_{X'} \tl =
0$ reduces to 
\bea
B d^2 \lambda + 2 ({d}^a B - B \tom^a ){d}_a \lambda 
  =  \alpha A \delta_\l \tl  \, \, . 
\label{Bint}
\eea
Thus, if $\alpha \leq 0$ this equation implies that 
\bea
B d^2 \lambda + 2 ({d}^a B - B \tom^a ){d}_a \lambda \geq 0 \, . \label
{al0}
\eea

Now, $\lambda$ must achieve a maximum on $S$, so we can define a 
new
function $\lambda^\star = \lambda - \lambda_{max}$ that also satisfies
equation (\ref{al0}) and achieves a maximum value of zero. Then, by 
our
usual maximum principle, since $\lambda^\star$ is not everywhere
negative, it must be constant. This means that $\lambda$ is also
constant and so by (\ref{Bint}), $\alpha = 0$.  Thus $\alpha \leq 0
\Rightarrow X'^a = \lambda_o X^a$ for some constant $\lambda_o$.
Similar reasoning shows that $\alpha \geq 0$ also implies that $X'^a$ 
is
a constant multiple of $X^a$.  $\Box$

%
\section{Horizons and their properties}
\label{FOTHs}
%
In this section we apply the properties of FOTS to gain a better
understanding of future outer trapping horizons.  First though we recall
some definitions. 

\subsection{Horizons}
\label{horizons}

We begin with the definition of a future outer trapping horizon (FOTH).
A \emph{trapping horizon} is a three-dimensional submanifold of a
spacetime $(M,g_{ab})$ that may be foliated with closed and spacelike
two-surfaces $S_v$ (where $v$ is a foliation parameter) on which $\tl 
=
0$ \cite{hayward}. Trapping horizons are classified by the values taken
by  $\tn$ and $\delta_n \tl$ on their leaves. A trapping horizon is said
to be future (past)  if $\tn < 0$ ($\tn >0$) while it is outer (inner)
if there is a scaling of the null vectors such that $\delta_n \tl <0$
($\delta_n \tl > 0$).%
\footnote{The original definition of \cite{hayward} is phrased in terms
of a dual-null foliation of the spacetime in some vicinity of the
horizon instead of the variations that we use here. In constructing such
a foliation, one must usually abandon the $\ell \cdot n = -1$
normalization \cite{HaywardDualNull} , however having done this the
definition can be phrased in terms of Lie derivatives rather than
variation operators.  That said, the definitions are equivalent.}
  These names are taken from the horizons that satisfy these 
conditions
in a fully extended Kerr or Reissner-Nordstr\"{o}m spacetime. Thus the
event horizon is a \emph{future outer trapping horizon (FOTH)} and 
the
inner Cauchy horizon is future inner. The corresponding white hole
horizons are past outer and past inner respectively.  In this paper, we
will be mainly in interested in FOTHs (which are foliated by FOTS).  

As noted in the introduction, apart from (future outer) trapping
horizons, there are other closely related quasilocal horizons. Here we
recall the definition of isolated and dynamical horizons, while in
section \ref{SEH} we will consider slowly evolving horizons. 

A three-dimensional submanifold of a spacetime $(M,g_{ab})$ is a
\emph{non-expanding horizon} if: i) it is null and topologically $S
\times \Rbar$ for some closed two-manifold $S$, ii) $\tl = 0$ and iii)
$-T^{ab} \ell_b$ is future directed and causal \cite{isoMech}. As usual
$\ell^a$ is an outward pointing normal and we note that since the
horizon is null, no foliation is required for its construction. However,
a foliation is certainly no hindrance to a three-surface being a
non-expanding horizon, and any null FOTH satisfying the null energy
condition will certainly be a non-expanding horizon. 

Non-expanding horizons are the simplest objects in the isolated 
horizon
family. Any non-expanding horizon can be turned into a \emph{weakly
isolated} horizon if the scaling of the null vectors is chosen so that
$\Lie_\ell \omega_a = 0$ for
\bea
\omega_a := - n_b \nabla_{\underleftarrow{a}} \ell^b 
  = -\kappa_{\ell} n_a + \tom_a \, , 
\eea
where the arrow indicates a pull-back into the cotangent bundle of the
non-expanding horizon. With this scaling zeroth and first laws of
isolated horizon mechanics may be established \cite{isoMech}.
Furthermore, an \emph{isolated} horizon is obtained by strengthening 
the
above conditions to require that the entire extrinsic geometry encoded
in the derivative operator be time independent.

Finally, a three-dimensional sub-manifold of a spacetime $(M,g_{ab})$ 
is
a \emph{dynamical horizon} if it: i) is spacelike and ii) can be
foliated by spacelike two-surfaces such that the null normals to those
surfaces satisfy $\tl = 0$ and $\tn < 0$ \cite{ak}. As we shall see in the 
next subsection, 
if the null energy condition holds and $\delta_\ell \tl \neq 0$ then
dynamical horizons are FOTHs and vice versa.

Non-expanding, isolated and dynamical horizons each have a rich set 
of
properties that may be derived directly from their definitions. However,
examples of spacetimes exist that contain (and are actually foliated 
by)
isolated horizons but do not contain trapped surfaces \cite{LewIso}.
Similarly there are spacetimes with dynamical horizons but no trapped
surfaces \cite{pp}. Thus if we take trapped surfaces as the defining
property of black holes, neither of these definitions is sufficient to
specifically single out black holes. Instead they represent necessary
conditions which must be supplemented to become sufficient. 

Given this observation, we will take FOTHs as our basic objects and
classify them by hybridizing the  naming systems. Thus a FOTH that 
also
satisfies one of these other sets of properties will be referred to as
non-expanding, weakly isolated, or dynamical as appropriate. 

%
\subsection{Properties of FOTHs}

A FOTH can be thought of as the variation surface associated with a
(finite) deformation of a FOTS (like the one shown in
Fig.~\ref{VarSurf}). Specifically given a FOTH and a foliation labelling
$v$, we can always find a tangent vector field $\mathcal{V}$ that is
normal to the $S_v$ and which satisfies 

\bea
\Lie_{\cV} v = 1 \, \, .  
\eea
Then, $\mathcal{V}^a$ can be viewed as a variation vector field and 
as
for other deformation vectors we write
\bea
\mathcal{V}^a = A \ell^a - B n^a \,  ,  
\eea
for some functions $A$ and $B$. Further, identifying points on the
different $S_v$  by the flow generated by $\mathcal{V}$, we write
derivatives of the two-geometry with respect to $v$ as 
\bea
\frac{d}{dv} \equiv \delta_\mathcal{V} \, \, . 
\eea
Then we may apply our results on FOTS to learn about FOTHs. 

\newtheorem{Hprop}{FOTH Property}

First, their topology is strongly constrained \cite{hayward} and this
follows directly from FOTS Property \ref{FOTStop}:

\begin{Hprop}
Let $H$ be a FOTH and assume the dominant energy condition. Then 
$H$ has
topology $S^2 \times \Rbar$. 
\end{Hprop}

Next, we consider the circumstances under which a FOTH is non-
expanding
and those by which it is dynamical.  In particular we will be interested
in transitions between these behaviours and so as a preliminary we
define non-expanding and dynamical sections of a FOTH. If a FOTH 
$H$ is
null (that is $B=0$ everywhere) for some range $v_1 \leq v \leq v_2$ 
of
the foliation parameter then we will refer to this as a non-expanding
section of $H$. In contrast if $H$ is spacelike (that is $AB > 0$
everywhere) for some range $v'_1 < v < v'_2$ of the foliation 
parameter
then we will refer to this as a dynamical section of $H$.

Then by FOTS Properties \ref{Adef0} and \ref{dynamic}, on any $S_v$ 
of a
FOTH either $B=0$ everywhere or $B \neq 0$ anywhere. Thus, no 
element
of the foliation is partly non-expanding and partly dynamical.
Transitions between non-expanding and dynamical sections must 
happen
``all-at-once".  \begin{Hprop} \label{transition} Let $H$ be a FOTH with
foliation $S_v$ and assume the null energy condition. Then $H$ may 
be
completely partitioned into non-expanding and dynamical sections. On
non-expanding sections $\delta_\l \tl = 0$ everywhere while on 
dynamical
sections $\delta_\l \tl < 0$ at least somewhere on each $S_v$.
\end{Hprop} 

This surprising result was first shown (using slightly different
assumptions) in \cite{ams}.  As one application, it
guarantees that a FOTH on which $\delta_\l \tl \neq 0$ 
for at least one point on each cross section is necessarily a 
dynamical horizon. The converse is slightly more restricted and 
we must require that $\delta_\ell \tl \neq 0$ everywhere to ensure that
a dynamical horizon also be a FOTH. 
 To see this consider that a dynamical horizon
is necessarily spacelike and so $X^a$ can always be oriented so that
$A>0$ and $B>0$. Scaling the null vectors so that $B$ is constant
on each cross-section of the horizon (\ref{Lvt}) simplifies to
\begin{equation}
  B \delta_n \tl = A \delta_{\ell} \tl \, \, . 
\end{equation}
Thus, if $\delta_{\ell} \tl \neq 0$ everywhere on a cross-section then the null energy condition guarantees that is strictly negative 
and so $\delta_n \tl < 0$ everywhere 
as well.  Consequently such a dynamical horizon is a FOTH. 
However, if $\delta_\ell \tl = 0$ somewhere 
(or everywhere) on the surface, then at those points $\delta_n \tl$ will also vanish
and the dynamical horizon will not be a FOTH. Examples
of such dynamical horizons are considered in \cite{pp}. 

Next by FOTS Properties \ref{FOTSgeom} and \ref{Dyngeom}, 
dynamical
versus non-expanding sections differ in more than just signature:

\begin{Hprop}\label{area_increase}

Let $\Delta H$ be a (section of a) FOTH and assume that the null 
energy
condition holds.

\begin{enumerate}

\item If $\Delta H$ is non-expanding, then the intrinsic geometry of the
$S_v$ is invariant. That is $\delta_{\cV} \tq_{ab} = 0$.  

\item If $\Delta H$ is dynamical, then $S_v$ locally increases in area
everywhere. That is $\delta_ {\cV} \vS = F \vS$ for some function $F$
that is everywhere positive. 

\end{enumerate}
\end{Hprop}

The first part of this property is well known from the isolated horizon
literature \cite{isoMech, isoGeom}.  The second part is essentially
Hayward's second law of black hole mechanics \cite{hayward}. FOTHs
expand if and only if they are dynamical. Otherwise their intrinsic
geometry is unchanging. 

Having reaffirmed these basic results, we can consider how FOTHs 
may be
deformed while preserving their defining characteristics. First we
consider variations of the foliation that leave $H$ itself invariant. 

\begin{Hprop}
Let $H$ be a FOTH and assume that the null energy condition holds. 

\begin{enumerate}

\item If $H$ is non-expanding then the foliation may be smoothly
deformed by any vector field of the form $X^a = f \ell^a$ where $f$ is
any positive function.

\item If $H$ is dynamical then the foliation is rigid and can be 
relabelled
but not deformed.  
\end{enumerate} 
\end{Hprop}

The first part of this property follows directly from FOTS Property
\ref{Adef0} which we apply to all of the $S_v$ simultaneously%
\footnote{Such variations of the foliation will generally only be
allowed for a finite range of the variation. Beyond that range we may
violate one of the defining conditions, for though the intrinsic
geometry of $H$ will remain invariant, the extrinsic geometry of the
two-surfaces normal to $H$ will change. Thus, finite variations may
ultimately generate foliation slices with $\tn$ or $\delta_n \tl$
non-negative.}%
.  The second part follows from FOTS Property \ref{TwoVarsProp} 
which
tells us that if $\delta_{\cV} \tl = 0$ then $\delta_{(\alpha \cV)} \tl
= 0$ if and only if $\alpha$ is a constant over $S_v $. For such a
constant $\alpha$ the variation would preserve (but relabel) the
foliation. 

Thus, the foliation may be deformed in an infinite number of ways on a
non-expanding (section of a) FOTH while it may not be deformed at all 
on
a dynamical section. The first part of this property is consistent with
the fact that particular foliations are not important for most of the
isolated horizon formalism.  The second part is consistent with the 
more
general result of \cite{gregabhay} which says that if $S$ is a FOTS 
and
$H$ is a dynamical FOTH with foliation $S_v$, then $S \subset H$ if 
and
only if it $S=S_v$ for some $v$. 

Next, we consider the other ways in which a FOTH may be deformed. 
Again
the non-expanding and dynamical cases are quite different. 

\begin{Hprop} \label{FOTHDef}
Let $H$ be a FOTH and assume the null energy condition. We 
consider
variations that deform $H$ but preserve the FOTH conditions.

\begin{enumerate}

\item If $H$ is non-expanding then all allowed variations map $H$ into
itself. 

\item If $H$ is dynamical then all allowed variations deform each leaf
of its foliation partly into the causal future and partly into the
causal past of $H$. 

\end{enumerate}
\end{Hprop}

This property also follows directly from FOTS Properties \ref{Adef0} 
and
\ref{TwoVarsProp} and can be rephrased to say that non-expanding 
FOTHs
are stable against deformations but dynamical FOTHs are not stable in
his way. The second part is consistent with \cite{gregabhay} where it
was shown that if $H_1$ and $H_2$ are dynamical FOTHs with 
foliations
$S_{v_1}$ and $S_{v_2}$ respectively, then no $S_{v_1}$ can lie 
entirely
in the causal past (or future) of $H_2$ and vice versa. Thus, either
$S_{v_1}$ is causally disconnected from $H_2$ or it intersects it so
that it lies partly in the causal future and partly in the causal past.
Property \ref{FOTHDef} may be thought of as a local version of that
global result. 

%
\section{Slowly evolving horizons and their flux laws}
\label{SEH}
%

A significant part of  physics deals with systems that are at or near
equilibrium.  For horizons, we naturally take non-expanding FOTHs to 
be
equilibrium states. Included in this class are the event horizons of all
(non-extreme) Kerr-Newmann black holes (whose properties are 
summarized
in Appendix \ref{KerrFOTS}).  Further, such FOTHs are automatically
isolated horizons and so all of the results from that formalism apply to
them. Recall too that by FOTH Property \ref{FOTHDef}, these surfaces
(locally) can only be deformed into themselves, so there is no 
ambiguity
about their exact location. 

It is then natural to consider near-equilibrium horizons, which should
be those that change slowly in time. Now, while this is a very
reasonable condition intuitively,  it is not so easy to geometrically
and invariantly characterize such horizons --- keep in mind that
dynamical FOTHs are spacelike and so there is no natural notion of 
time
intrinsically associated with the surface. In this section we will
motivate a definition of these \emph{slowly evolving horizons} and 
then
explore some of its consequences. A version of the definition and 
many
of the properties appeared in \cite{prl} but most of the derivations
appear here for the first time. 

\subsection{Slowly expanding horizons}

Intuitively, we would expect the properties of a slowly expanding
horizon to be ``close'' to those of a non-expanding or isolated horizon.
In this section we will formalise this requirement.  This is essentially
done by performing a perturbation expansion around a non-expanding
horizon in powers of a small parameter $\epsilon$.  However, as we 
shall
see, care must be taken as spacelike and null surfaces are 
fundamentally
different, and in particular it is non-trivial to find a normalization
which transitions smoothly between the dynamical and isolated 
regimes.

In order to proceed, we will slightly modify the formalism introduced
over the last three sections.  First we loosen it so that while the
evolution vector field $\cV^a$ still generates the foliation, it is no
longer tied to a specific labelling; that is we only require that
$\Lie_{\cV} v = \alpha(v)$ for some positive function $\alpha(v)$.  Next
we tighten it by choosing $A=1$ so that 
\bea
\cV^a  = \ell^a - C n^a \, ,  \label{cV}
\eea
where we change notation from $B$ to $C$ to avoid confusion with the
more general variations considered previously.  The net effect is to
reduce the scaling freedom of the null vectors to that contained in
$\alpha(v)$ so that for a fixed foliation labelling
\bea
\cV^a = \alpha \cV_o^a \, \, , 
\ell^a = \alpha \ell_o^a \, \, , 
n^a = \frac{1}{\alpha} n_o^a \, \, \mbox{and} \, \, 
C = \alpha^2 C_o \, ,  \label{rescaling_freedom}
\eea
where the subscript ``o'' indicates a quantity defined with $\alpha = 1$. 

If a FOTH is null and non-expanding, then $C=0$ and the evolution 
vector
field $\cV = \ell$.  Thus, the most direct way to define a slowly
expanding horizon might seem to be to say that it is a (section of) a
horizon on which $C$ is very small.  Unfortunately this is not a viable
strategy as one can use the rescaling freedom (\ref
{rescaling_freedom})
to make $C$ arbitrarily small on any $\triangle H$.  Instead we 
proceed
by focusing on how the $S_v$ are evolved by $\cV^a$. 

For our purposes non-expanding horizons have two key properties: 
they
are null and the intrinsic geometry of their foliating two-surfaces is
invariant. Our invariant characterization of a slowly expanding horizon
draws on both of these ideas. First, we consider the evolution of the
area-form $\vS$ on $S_v$. From (\ref{LcV}) we have
\bea
\Lie_{\cV} \vS = - C \tn \vS \, , 
\eea
which is certainly scaling dependent. However this dependence may 
be
easily isolated by rewriting
\bea
\Lie_{\cV} \vS = - C \tn \vS = || \cV ||  
\left(- \sqrt{\frac{C}{2}} \tn \vS \right) \, ,
\eea
so that the rescaling freedom is restricted to $||\cV||$. The term in
parentheses then provides an invariant measure of the rate of 
expansion.
Among other properties it vanishes if the horizon is non-expanding and
on a dynamical horizon section is equal to the rate expansion of $\vS$
with respect to the unit-normalized version of the evolution vector
field. We will consider the rate of expansion to be slow if this is
small and this approach is borne out by the examples of \cite
{billsPaper}.
In particular if $-\sqrt{C} \tn$ is of order  $\epsilon$ and the scaling
of the null vectors is chosen so that $|| \cV ||$ is commensurate (as
would be reasonable for an ``almost-null" surface) then $\Lie_{\cV} \vS
$
will be of order $\epsilon^2$. 

This notion of a slow area change captures the essence of a slowly
expanding horizon however we still need to require both that the 
surface
be ``almost-null" and that the rest of the intrinsic geometry also be
slowly changing. We have already seen that restrictions on the norm of
$\cV^a$ are not invariant, so instead we will implement these ideas
together by requiring that the evolution of the two-metric be
characterized by the expansion and shear associated with $\ell^a$ (as
they would be for a truly null surface). That is from (\ref{Lq}), 
\begin{equation}\label{Ltq_sig}
  \Lie_{\cV} \tq_{ab} = 
 2  \sigma^{(\ell)}_{ab}
  + \left( -  C \tn \tq_{ab} - 2 C\sigma^{(n)}_{ab} \right) 
  \approx  2 \sigma^{(\ell)}_{ab} \,  .  \nn \\
\end{equation}
Further, the matter flow across the horizon similarly should be, to
lowest order, the same as that across a null surface. That is we expect
\bea
\cV^a T_a^b \tau_b \approx  T_{ab} \ell^a \ell^b 
\, , \label{MatFlux}
\eea
where $\tau_a = \ell_a + C n_a$ is the usual timelike normal.  Given
that these quantities vanish on a non-expanding horizon we would 
also
expect both of these quantities to be small. An invariant 
implementation
of these ideas gives rise to the following definition:

\smallskip
\noindent
{\bf Definition:} Let $\triangle H$ be a section of a future outer
trapping horizon foliated by spacelike two-surfaces $S_v$ so that
$\triangle H = \{ \cup_v S_v : v_1 \leq v \leq v_2\}$. Further let
$\cV^a$ be an evolution vector field that generates the foliation so
that $\Lie_{\cV} v = \alpha(v)$ for some positive function $\alpha(v)$, and
scale the null vectors so that $\cV^a = \ell^a - C n^a$. Then $\triangle
H$ is a \emph{slowly expanding horizon} if the dominant energy 
condition
holds and

\begin{enumerate}

\item $\epsilon \ll 1$ where 
\begin{equation}\label{epsilon}
  \epsilon^2/R_H^2 =  \mbox{Maximum} 
  \left[C \left( \norm\sigma^{(n)}\norm^2 
  + T_{ab} n^a n^b + \tn^2/2 \right) \right]
  \, ,
\end{equation}

\item $\displaystyle |\tilde{R}|, \norm \tom \norm^{2}$ and $T_{ab}
\ell^a n^b \lesssim 1/ R_H^2$, where $R_{H}$ is the areal radius
 $\sqrt{a/4\pi}$ of the horizon, and

\item two-surface derivatives of horizon fields are at most of the same
order in $\epsilon$ as the (maximum of the) original fields.  For
example, $\displaystyle \norm d_a C\norm \lesssim C_{max}/R_H$, 
where
$C_{max}$ is largest absolute value attained by $C$ on $S_v$. 

\end{enumerate}
Throughout this definition and what follows, an expression like $X
\lesssim Y$ means that $X \leq k_o Y$ for some constant $k_o$ of 
order
one, and in particular $k_o \ll 1/\epsilon$.

As we will shortly see, the first condition is simply a scaling
invariant way of writing our earlier requirements that the horizon
geometry be slowly changing and that those changes be dominated by 
the
$\ell$-components of quantities.  Additionally, it introduces a simple
way to define a value for $\epsilon$ on each cross section of the
horizon.  If the $\alpha(v)$ is chosen so that 
\begin{equation}\label{Cscale}
  C \approx \epsilon^2 \, ,
\end{equation}
then this scaling of the null normals is compatible with the
\emph{evolution parameter} $\epsilon$. 

The second and third conditions are restrictions on the horizon
geometry.  Effectively, they ensure that the geometry of the horizon is
not too extreme.  These conditions are rather mild, and as a partial
justification for these assumptions in Appendix \ref{KerrFOTS} it is
shown that they all hold on a Kerr horizon with the standard foliation.

We now consider the implications of these assumptions. To begin, 
by equation (\ref{dntl}) our restrictions on the horizon geometry
immediately imply that
\bea
|\delta_n \tl |  \lesssim \frac{1}{R_H^2} \, .
\eea
Note the use of the absolute value sign. Even though we must have
$\delta_n \tl < 0$ for some scaling, in general this need not be true
for the particular $A=1$ scaling that we have chosen.  

Next, we can bound the flux of matter and the gravitation shear at the
horizon.  To do this, we make use of the fact that $\delta_{\cV} \tl =
0$.  Then, equation (\ref{Lvt}) along with the magnitude of $C$ fixed 
by
(\ref{Cscale}) implies that 
\bea
| \delta_\ell \tl | = \norm\sigma^{(\ell)}\norm^2 
  +  8 \pi G T_{ab} \ell^a \ell^b \lesssim \frac{\epsilon^{2}}{R_H^2} \, ,   
\label{sllTll}
\eea
or, using the null energy condition,
\bea
\norm\sigma^{(\ell)}\norm^2 \lesssim \frac{\epsilon^{2}}{R_H^2} 
\, \, \, \mbox{and} \, \, \, 
8 \pi G T_{ab} \ell^a \ell^b \lesssim \frac{\epsilon^{2}}{R_H^2} \, \, . 
\label{drive_terms}
\eea
Thus these terms are bound by the size of $\epsilon$.  Explicit 
examples
of these bounds in action may be found in \cite{billsPaper} where both
matter and shear driven expansions are considered in some detail. 

From these results it follows that the two-metric is slowly changing
with the highest order contributions coming from quantities associated
with $\ell^a$. We have 
\bea
\Lie_{\cV} \tq_{ab} = 
   \underbrace{2 \sigma^{(\ell)}_{ab}}_{O(\epsilon)} 
  + \underbrace{\left( C \tn \tq_{ab} - 2 C\sigma^{(n)}_{ab}
\right)}_
  {O(\epsilon^2)} \, , \label{LcVex}
\eea
while
\bea
\Lie_{\cV} \vS = \underbrace{- C \tn \vS}_{O(\epsilon^2)} \, .
\eea
In addition, there is a bound on the rate of change of the intrinsic
curvature.  The evolution of $\tilde{R}$ is derived in (\ref{2curv_var})
and is given by
\begin{equation}
  \Lie_{\cV} \tilde{R} = 
  2 d^{a} d^{b} (\sigma^{(\ell)}_{ab} - C \sigma^{(n)}_{ab}) 
  + d^{2} ( C \theta_{(n)}) 
  + C \theta_{(n)} \tilde{R} 
  \, .
\end{equation} 
From this we see immediately that $\Lie_{\cV} \tilde{R} \lesssim
\epsilon/R_{H}^{3}$.  Thus, the intrinsic geometry of the horizon is
slowly varying, at a rate $\epsilon$.  The choice (\ref{Cscale})
effectively scales $\cV^a$ to reflect the slowly expanding nature of the
horizon.  In a transition to isolation so that $\epsilon \rightarrow 0$
this choice will force $\norm \cV \norm \rightarrow 0$, thus ensuring 
that 
the limit is continuous.

We can also bound the flux of energy momentum through the horizon.  
We
have already seen that $T_{ab}\ell^{a}\ell^{b} \lesssim
\epsilon^{2}/R_{H}^{2}$.  Furthermore, this is the main flux of
energy through the horizon as our definition implies that
\begin{equation}
  8 \pi G \cV^a T_a^b \tau_b = 
  \underbrace{8 \pi G T_{ab} \ell^a \ell^b}_{O(\epsilon^2)} 
  -  \underbrace{8 \pi G C^2 T_{ab} n^a n^b}_{O(\epsilon^4)} \, \, . 
  \label{VTtex}
\end{equation}
If the dominant energy condition holds, then there is a further
constraint on the components of the stress-energy tensor.  In that 
case,
with $\tau^a$ future-directed and timelike on a dynamical FOTH, $-
T_{ab} \tau^b$ must also be future directed and causal. Then, 
\bea
g^{ab} (T_{ac} \tau^c) (T_{bd} \tau^d) \leq 0 
\; \;  \Rightarrow \; \;  
\norm\tq_a^b T_{bc} \tau^c \norm^2 \leq  
  2 (T_{ab} \ell^a \tau^b)(T_{cd} n^c \tau^d) \, \, .  
\eea
However, by (\ref{epsilon}), (\ref{Cscale}) and (\ref{drive_terms}),
$(T_{ab} \ell^a \tau^b)(T_{cd} n^c \tau^d)$ is of order
$\epsilon^2/R_H^4$. Thus, 
\bea
\norm \tq_a^b T_{bc} \tau^c \norm \lesssim \frac{\epsilon}{R_H^2} 
\, \, . \label{GenAngFlux}
\eea
This result, in conjunction with (\ref{expl2deriv}) can be used to bound
one of the components of the Weyl tensor.  We obtain 
\begin{equation}
  \tq_{a}^{b} C_{bcde} \ell^{c} \ell^{d} n^{e} \lesssim
  \frac{\epsilon}{R_{H}^{2}} \, .
\end{equation}
This is equivalent to $\Psi_{1} = C_{abcd}\ell^{a} m^{b} \ell^{c} n^{d}
\lesssim \epsilon/R_{H}^{2}$.  For an isolated horizon the equivalent
quantities vanish, namely $\tq_a^b T_{bc} \ell^c =0$ and $\Psi_{1} = 0
$
\cite{isoMech}.  

The flux of incoming gravitational radiation is encoded in $\Psi_{0} =
C_{abcd} \ell^{a} m^{b} \ell^{c} m^{d}$, another of the components of
the Weyl tensor.  For a horizon to be slowly expanding, one would 
expect
this quantity to be small.  We can show that this is the case by 
considering
the evolution of the shear $\sigma^{(\ell)}$, derived in (\ref{siglVar}).
Keeping only the lowest order terms, we obtain
\begin{equation}
  \Lie_{\cV} \sigma^{(\ell)}_{ab} - \kappa_{\cV} \sigma^{(\ell)}_{ab} =
  \tq_a^c \ell^d \tq_b^e \ell^f C_{cdef} + O(\epsilon^{2}) \, .
\end{equation}
Therefore $\sigma^{(\ell)}$ will remain small only if $\Psi_{0}
\lesssim \epsilon/R_{H}^{2}$.

Finally, we turn to $\Psi_{2} = C_{abcd} \ell^{a} m^{b} \bar{m}^{c}
n^{d}$.  Begin by noting that $\Psi_{2}$ is invariant under rescalings
of $\ell$ and $n$.  On an isolated horizon, the value of $\Psi_{2}$ is
not restricted, however, $\Lie_{\mathcal{V}} \Psi_{2} = 0$.  For a
vacuum, slowly evolving horizon, $\Lie_{\mathcal{V}} \Psi_{2} \lesssim
\epsilon / R_{H}^{3}$.  This follows from the Gauss and Ricci relations 
of Appendix
\ref{Prel} which can be used to rewrite $\Psi_{2}$ in terms of the intrinsic
and extrinsic horizon geometry. The result then follows directly
from the fact that $\norm \sigma^{(\ell)} \norm \lesssim \epsilon /
R_{H}$ and the fact that $\tom$, $\tilde{R}$ and $\sigma^{(\ell)}$ are
all slowly evolving.

To summarize, we have seen that the definition of a slowly expanding
horizon captures many expected properties of a near equilibrium black
hole.  Specifically, the intrinsic geometry, including the area and
two-curvature is slowly changing and there is little flux of matter or
gravitational energy through the horizon.  Of course, the given orders
of quantities are bounds rather than requirements on the size of those
terms.  For example, on a spherically symmetric slowly expanding
horizon, $\sigma_{ab}^{(\ell)}$ vanishes identically and so the metric
is unchanging at first order. Similarly in a vacuum spacetime where a
horizon grows through the absorption of gravitational waves, all matter
terms will vanish. Examples of both of these behaviours may be found in \cite{billsPaper}.

\subsection{Slowly evolving horizons and the first law}

In the previous section, we have placed requirements on the intrinsic
geometry of the horizon and arrived at the notion of a slowly 
expanding
horizon.  Now, we shall impose some further restrictions in order to
obtain the first law of black hole mechanics.  This will be done by
restricting the extrinsic geometry of the horizon.  The inspiration for
the extra conditions comes from the weakly isolated horizons defined 
in
section \ref{horizons}.  A non-expanding horizon becomes weakly 
isolated
if the scaling of the null vectors is chosen so that 
\bea
\Lie_\ell \omega_a = 0 \, \, .  \label{Llom}
\eea
In this case it is also true \cite{isoGeom} that one can always find a
``good" foliation of surfaces $S_v$ such that 
\bea
\Lie_\ell \tn = 0 \, \, . \label{LtnIso}
\eea
With this foliation, a suitable scaling sets $n_a = - [dv]_a$ and
(\ref{Llom}) can be decomposed as
\bea
\Lie_\ell \kappa_{\ell} = 0 
\; \; \mbox{and} \; \; 
\Lie_\ell \tom_a = 0 \, \, . 
\eea
We will enforce versions of these conditions to obtain slowly evolving
horizons.  However, there is a distinction between the slowly evolving
and isolated cases.  Since non-expanding horizons do not have a
pre-determined, fixed foliation, it is \emph{always} possible to rescale
the null normal $\ell$ so that conditions (\ref{Llom}) and
(\ref{LtnIso}) are satisfied.  In contrast, a non-equilibrium horizon
comes endowed with a unique foliation, so the equivalent conditions 
are
not guaranteed to hold, they will have to be checked.  This motivation
leads us to the following:
\smallskip

\noindent
{\bf Definition:} Let $\triangle H$ be a slowly expanding section of a
FOTH with a compatible scaling of the null normals. Then it is said to
be a \emph{slowly evolving horizon} if in addition

\begin{enumerate}

\item $\displaystyle \norm\Lie_{\cV} \tom_a \norm$ and $|\Lie_{\cV}
\kappa_{\cV}|  \lesssim \epsilon / R_H^2$ and 

\item $|\Lie_{\cV} \tn| \lesssim \epsilon/R_H^2$. 

\end{enumerate}

The first consequence of the above definition is that on a slowly
evolving horizon, the surface gravity is slowly varying.  It follows
immediately from the definition that $\kappa_{\cV}$ is slowly changing
in time.%
\footnote{As for a weakly isolated horizon \cite{isoMech} this condition
can equivalently be implemented by imposing conditions on the 
various
quantities that arise if one takes the  $\Lie_{\cV}$ derivative of
equation (\ref{expnderiv}) and then deriving the desired result for
$\Lie_{\cV} \kappa_{\cV}$.  However for simplicity we just impose the
condition directly.}%
  The fact that it is nearly constant across each two-surface follows
from (\ref{ang_mom_expression}).  Keeping only the lowest order 
terms,
we have:
\begin{equation}
  \Lie_{\cV} \tom_{a} = d_{a} \kappa_{\cV} - d^{b} \sigma^{(\ell)}_{ab}
  - 8 \pi G \tq_{a}^{b} T_{bc} \ell^c + O(\epsilon^{2}) \, .
\end{equation}
From the definition above, (\ref{drive_terms}) and (\ref{GenAngFlux}),
it follows immediately that
\begin{equation}\label{dkappa} 
  \norm d_a \kappa_{\cV} \norm \lesssim 
  \frac{\epsilon}{R^2_H} \, \, . 
\end{equation}
That is, the surface gravity is approximately constant over each slice
of the foliation.  Since we have also required the surface gravity to be
slowly changing up the horizon, it follows that over a foliation
parameter range on $\triangle H$ that is small relative to $1/\epsilon$, 
\begin{equation} \label{kap0}
  \kappa_{\cV} = \kappa_o + {O}(\epsilon)
\end{equation}
for some constant $\kappa_o$. Note however, that if a FOTH is slowly
evolving for long enough, then larger changes can accumulate. 

Slowly evolving horizons obey a first law of black hole mechanics
\cite{prl}. Applying the slowly evolving horizon conditions to equation
(\ref{constraint_non_rot}), it reduces to a dynamical version of this
first law. To order $\epsilon^2$ we obtain: 
\begin{equation}\label{FirstLaw}
  \frac{\kappa_o \dot{a}}{8 \pi G} \approx 
  \int_{S_v} \vS \left\{\frac{\norm\sigma^{(\ell)}\norm^2}{8 \pi G} 
  + T_{ab} \ell^a \ell^b \right\} \, , 
\end{equation}
where $a$ is the area of the two-surfaces and $\dot{a} = \int_{S_v}
\Lie_{\cV} \vS$.  Interestingly, in \cite{Gourgoulhon:2006uc}, equation
(\ref{constraint_non_rot}) has been interpreted as a second order
evolution equation for the horizon area.  In the slowly evolving limit,
this reduces to (\ref{FirstLaw}).  In section \ref{OtherFlux} we will
compare this version of the first law other well-known flux laws, but
for now consider it in its own right. 

Let us examine the two energy flux terms contributing to the area
increase.  The first is the square of the gravitational shear at the
horizon, while the second is the flux of matter stress energy through
the horizon.  The first term is necessarily positive, and provided the
null energy condition holds, so is the second.  Then, FOTH Property
\ref{area_increase} implies that for a dynamical slowly evolving 
horizon
$\kappa_o > 0$ -- the average surface gravity of a slowly evolving
horizon is necessarily positive. See \cite {bfExtreme} for further
discussion of this point and its relation to extremal horizons. 

Next, we would like to examine whether (\ref{FirstLaw}) can be
integrated to give a value for the horizon energy.  For a slowly
evolving horizon, (\ref{expnderiv}) reduces at leading order to 
\bea
  \kappa_o \tn  = 
  - \tilde{R}/2 + 8 \pi G T_{ab} \ell^{a} n^{b} 
  + \norm\tilde{\omega}\norm^2   + {d}_{a} \tilde{\omega}^{a} 
  + \mbox{O}(\epsilon)   \, \, . 
\eea
For a spacetime which is close to spherically symmetric, such as
those considered in \cite{billsPaper}, the $ \tom_a$ terms are of order
$\epsilon$ or smaller while if the only matter is radially infalling
dust the matter term may also be neglected. Then 
\bea
  \kappa_o \tn \approx - \frac{\tilde{R}}{2} = - \frac{1}{R_H^2} \, \, . 
\eea
If one scales the null vectors so that $\tn = - 2/R_H$, which is the
value taken in the Schwarzschild spacetime, it follows that $\kappa_o 
=
1/{2R_H}$.  Then it is immediate that (\ref{FirstLaw}) may be 
integrated
to give an energy of $E \approx R_H/2$.  In this case, the first law can
be written as 
\bea
  \dot{E} \approx  \frac{d}{dv}  \left(\frac{\kappa_o a}{4 \pi G} \right) 
  \approx \frac{\kappa_o \dot{a}}{8 \pi G}    
  \approx \int_{S_v} \vS \left\{\frac{\norm\sigma^{(\ell)}\norm^2}{8 \pi G} 
  + T_{ab} \ell^a \ell^b \right\} \, , 
\eea
where we have taken the foliation label $v$ to be compatible with null
scaling (that is $\Lie_{\cV} v = 1$).  Thus we recover all of
the standard notions of black hole mechanics: the energy is given by 
the
Smarr formula and its time rate of change may be written in terms of
both $\kappa_o \dot{a}$ and a flux law.  In more general situations
however, things are not quite so tidy.  While (\ref{FirstLaw}) always
holds, away from spherical symmetry and in the presence of 
alternative
matter fields the later simplifications cannot be made.  Thus, in general
it is not guaranteed that (\ref{FirstLaw}) will integrate to a tidy
expression for the energy -- this is not too surprising given the well-
known 
uncertainties in defining (quasi)localized gravitational energy. 

\subsection{Approximate symmetries and angular momentum}
\label{SEH_angMom} 

We would like to generalize the first law for slowly evolving black
holes to include angular momentum.  We begin by noting that on a 
slowly
evolving horizon,  (\ref{ang_mom_evolution}) simplifies to
\bea \label{LVJ_slow}
\dot{J}[\phi] \approx  \int_{S} \vS \left\{  
\frac{ \sigma^{(\ell)} \mspace{-6mu} : \mspace{-4mu} \sigma^{(\phi)}}{8 
\pi G} 
+ T_{ab} \ell^{a} \phi^
{b}  \right\}  \, ,
\eea
where both terms are $O(\epsilon)$ due to (\ref{drive_terms}) and
(\ref{GenAngFlux}) respectively.  Thus, the angular momentum 
associated
to \emph{any} rotation vector field $\phi$ must be slowly varying, even
if $\phi$ is not a symmetry of the horizon.  However, in this case the
change in the angular momentum is only restricted to be at most of 
order
$\epsilon$ while the area (and energy) evolve at a rate proportional to
$\epsilon^{2}$.  This reflects the fact that we have not required the
vector field $\phi$ to be a symmetry of the horizon.  Since the horizon
is not in equilibrium, we do not expect it to possess an exact 
symmetry,
and instead introduce the notion of an approximate symmetry.

\smallskip
\noindent

{\bf Definition:} Let $\triangle H$ be a section of a FOTH and $\phi^a
\in TS_v$ be a rotation vector field as defined in section \ref{angMom}.
Then $\phi^a$ is said to be an \emph{approximate symmetry} of the
horizon if:

\begin{enumerate}

\item $\displaystyle \norm \Lie_\phi \tq_{ab} \norm \lesssim
\epsilon/R_H^2$,   

\item $\displaystyle | \Lie_\phi \tn | \lesssim \epsilon/R_H^2$, and

\item $|\displaystyle 8 \pi G T_{ab} \phi^a \tau^b| \lesssim
\epsilon^2/R_H^2$. 

\end{enumerate}

The first two conditions require $\phi^{a}$ to be an approximate
symmetry of the intrinsic and (part of) the extrinsic geometry of the
two-surfaces. The third condition is not really a symmetry condition but
instead says that the angular matter flux should be particularly small
in the $\phi^a$ direction (in general (\ref {GenAngFlux}) only restricts
such fluxes to be of order $\epsilon$).  These conditions are sufficient
to guarantee that the angular momentum measured relative to $\phi^a $
changes to order $\epsilon^2$ with the expression for the rate of change
given as before by (\ref{LVJ_slow}).  Furthermore, the change in angular
momentum is proportional to a gravitational plus a matter flux. As for
other slowly evolving relations, these fluxes are calculated as if
$\triangle H$ was a null surface.  Finally, we note that the condition
$\Lie_{\cV} \phi^{a} = 0$ condition can be weakened to $\displaystyle
\norm \Lie_{\cV} \phi^a \norm \lesssim \epsilon^3/R_H^2$, without
affecting the angular momentum evolution.  This allows for slight
changes in the approximate symmetry direction as the horizon evolves. 

Now, let us consider the first law for slowly evolving, rotating
horizons.  In this case, we allow for a more general evolution vector
$\xi^a$, tangent to the horizon but with components both normal and
tangent to the cross sections.  We restrict the allowed $\xi^a$ in the 
following manner. First $\xi$ should preserve the
foliation of the horizon. That is $\xi^{a}= \cV^a + \Phi^a$, 
for some normalization of $\cV^{a}$ and some $\Phi^a$ which
is tangent to the horizon cross sections. Second, this $\Phi^a$
should generate rotations. That is, it should integrate to a flow which foliates the cross sections into two fixed points plus a congruence of 
closed curves and further those closed curves should have a common period (see \cite{theBeast} for a further discussion of rotations). 
Finally, we require that $\xi^{a}$ respect the slowly evolving nature 
of the horizon in a non-trivial way. That is we require that  
$||\xi|| \gg \epsilon$ while the norms of 
$\Lie_{\xi} \, \tilde{q}_{ab}$, $\Lie_{\xi} \, \tn$ 
and $\Lie_{\xi} \, \tom_a$ should be of order $\epsilon$.  Then, we can
write 
\bea
\xi^a = \cV^a + \Omega \phi^a \, ,
\eea
where $\phi^a$ is an approximate symmetry and $\Omega$ is an angular velocity which is constant on each horizon cross section. 
\footnote{Note that the angular velocity $\Omega$ is not related to the
curvature of the normal bundle ${\Omega}_{ab}$.}  
In addition, we require that $\Omega$ changes only very slowly
with respect to $\cV$ so that  $\Lie_\cV \Omega \lesssim
\epsilon^3/R_H^2$.  The evolution vector field $\xi$ is then the
analogue of the Killing evolution vector field on a Kerr horizon.  Thus,
combining (\ref{FirstLaw}) and (\ref {LVJ_slow}) and expanding to order
$\epsilon^{2}$ we obtain the first law for rotating horizons:
\bea
  \frac{\kappa_o \dot{a}}{8 \pi G} + \Omega \dot{J}[\phi] \approx 
  \int_{S} \vS \left\{  
  \frac{ \sigma^{(\ell)} \mspace{-6mu} : \mspace{-4mu} \sigma^{({\xi})} }
  {8 \pi G} 
  + T_{ab} \ell^{a} {\xi}^{b}  \right\}  \, \, .    
\label{FullFirstLaw}
\eea
To this order the fluxes can be calculated using $\xi^a \approx \ell^a +
\Omega \phi^a$. 

As in the last section this familiar form of the first law will hold for
all horizons of the considered class.  Again however, care must be 
taken
in the choice of scaling for the null vectors and choice of $\Omega$ if
one hopes to be able to integrate it to give a simple energy expression
for a rotating FOTH.  That said, for perturbations around a Kerr horizon
it is possible to scale the null vectors accordingly and so obtain the
standard functional dependence of energy on area and angular 
momentum.

\section{Comparison with other flux laws}
\label{OtherFlux}

The forms of the first law derived in the last section are not new. Other
dynamical flux laws may be found in the literature which apply to other
types of horizons. In this section we see how two of these laws may 
also
be derived from the deformation equations of section \ref{twoSurf} and
compare them to our form of the first law. 

\subsection{The first law for event horizons}
\label{EH}

A version of the first law usually known as the Hawking-Hartle formula
can also be derived for event horizons.  We now outline how it arises
with our chief focus being a comparison with the first law for slowly
evolving horizons.  For further details (and a slightly different
derivation) see the original paper \cite{hawkhartle}. 

An event horizon is a null surface ruled by a congruence of geodesics.
Thus its evolution is governed by Raychaudhuri's equation
\cite{hawkellis} which from our point of view is either equation
(\ref{constraint_non_rot}) or (\ref{Lvt}) with $A=1$ and $B=0$:
\begin{equation}\label{Ray}  
   \delta_\ell \theta_{(\ell)} = \kappa_\ell \tl 
       - \norm\sigma^{(\ell)}\norm^{2} 
     - G_{ab} \ell^{a} \ell^{b}
     -  (1/2) \theta_{(\ell)}^{2} \, \, .  
\end{equation}
While this is not strictly necessary for a null surface, we'll assume
that the horizon is foliated by spacelike two-surfaces and that the
foliation is compatible with affine scalings of the $\ell^a$.  Thus, we
could scale this null evolution vector so that $\kappa_\ell = 0$.
However, even if we don't, it is immediate that $\kappa_\ell$ is a
function of $v$ alone and so $d_a \kappa_\ell = 0$. 

Then, multiplying by the area element and writing variations in Lie
derivative form, we have
\bea
  \left(\kappa_{\ell} +  \tl/2 \right) \Lie_\ell \vS = 
  \Lie_{\ell} \left( \Lie_{\ell} \vS  \right) 
   + \vS \left(  \norm \sigma^{(\ell)} \norm^2 + G_{ab} \ell^a \ell^b \right )  
  \, \, .  
\eea
If we think of this as a perturbation of a standard (stationary) black
hole solution then the scaling of the null vectors should be such that
the surface gravity $\kappa_\ell$ is of order $1/R_H$ while other
quantities appearing in the above equation should be close to zero.
Thus we should have $\tl \ll \kappa_\ell$ and so can drop the $\tl/2$ on
the left-hand side of the above equation.  Then, integrating both over
$S_v$ and ``up" the horizon between two surfaces $S_1$ and $S_2$ 
we find
that  
\bea
  \int_{v_1}^{v_2}  \mspace{-12mu} dv  (\kappa_{\ell} \dot{a}) \approx 
  \left. \dot{a} \right|^{S_2}_{S_1}
  + \int_{v_1}^{v_2}  \mspace{-12mu} dv \int \vS 
  \left( \norm \sigma^{(\ell)} \norm^2 + G_{ab} \ell^a \ell^b \right ) \, . 
\label{EHev}
\eea
While it is certainly true that an event horizon always expands until it
reaches an ultimate equilibrium state, it is also true that during
periods of quiescence when nothing much is happening it can 
approach an
isolated horizon.  In particular, as discussed in \cite{meRev},
$\dot{a}$ can become arbitrarily close to zero.  If we consider an
evolution between two such ``equilibrium" states, the first term on the
right-hand side of (\ref{EHev}) can also be neglected.  Thus, we arrive
at the Hawking-Hartle formula:
\bea
\int_{v_1}^{v_2}  \mspace{-12mu} dv  (\kappa_{\ell} \dot{a}) \approx
\int_{v_1}^{v_2}  \mspace{-12mu} dv \int \vS 
\left(  \norm \sigma^{(\ell)} \norm^2 + G_{ab} \ell^a \ell^b \right ) \, . 
\label{EHev2}
\eea

This is the event horizon analogue of our (\ref{FirstLaw}). However, the
equation does not imply a causal relationship from fluxes to changes 
in
area.  This reflects the teleological nature of event horizons.
Expansions of event horizons are caused by an absence of 
interactions,
while fluxes through them instead force decreases of the rate of
expansion (again see \cite{meRev}). One of the significant implications
of the Hawking-Hartle formula is that time averages smooth out these
strange behaviours. 

Despite the very different character of FOTHs and event horizons, 
there
are remarkable similarities between the first laws for slowly evolving
horizons (\ref{FirstLaw}) and event horizons (\ref{EHev2}). In both
cases we must impose a condition which forces the horizon to be 
slowly
evolving in order to obtain the first law.  Thus, they both only hold
near equilibrium. Further, neither of these laws either specifies (or
requires) an energy definition for the black hole. 

If we further assume that the event horizon has an approximate 
symmetry
generated by a spacelike vector field $\phi^a$ so that $\Lie_\ell
\phi^a$ and $\Lie_\phi \tl$ can be neglected then
(\ref{ang_mom_evolution}) again becomes
\bea
  \dot{J}[\phi] \approx  
  \int_{S} \vS \left\{ 
  \frac{ \sigma^{(\ell)} \mspace{-6mu} : \mspace{-4mu} \sigma^{(\phi)} }
  {8 \pi G} 
  + T_{ab} \ell^{a} \phi^{b}  \right\}  \, \, . 
\label{LVJ_EH}
\eea
In contrast to (\ref{EHev2}) this is a snapshot rather than
time-integrated flux law.  This is a reflection of the much broader
applicability of the angular momentum flux law which holds not just for
horizons but for any surface \cite{by, ak, GourgFlux, fatherOfTheBeast,
theBeast, haywardRef}. 

Then, given a slowly varying angular velocity $\Omega$, we can 
define an
evolution vector field
\bea
\xi^a = \ell^a + \Omega \phi^a \, , 
\eea
and combine (\ref{EHev2}) and (\ref{LVJ_EH}), to get the more
general dynamical first law for event horizons
\bea
\int_{v_1}^{v_2} \mspace{-8mu} dv \left\{ \frac{\kappa_o \dot{a}}{8 \pi 
G} 
+ \Omega \dot{J}[\phi] \right\} \approx 
\int_{v_1}^{v_2}  \mspace{-8mu} dv \int_{S} \vS 
\left\{ \frac{ \sigma^{(\ell)} \mspace{-6mu} : \mspace{-4mu} \sigma^
{({\xi})} }
{8 \pi G} 
+ T_{ab} \ell^{a} {\xi}^{b}  \right\}  \, \, .    
\eea
Apart from the time-integration this is, of course, the same as the
corresponding law (\ref {FullFirstLaw}) for slowly evolving horizons.

\subsection{Dynamical horizon flux law}

Finally, we consider the dynamical horizon flux law of \cite{ak,
haywardFlux}.  Despite the by now familiar flux terms that appear in
this equation, it is different from the first laws that we have already
seen.  Instead of linking a $\kappa \dot{a}$ term to fluxes of
gravitational and matter stress energy through the horizon, it is
concerned with how these fluxes change the energy associated with 
the
black hole.  As we shall see, it is not clear how the approaches can be
directly connected. 

Let $H$ be a FOTH and further let $E(v)$ be any function that is
increasing whenever $H$ is dynamical but remains constant whenever 
(if
ever) it is isolated. The obvious choice is some function of the area
but in principle other any other function with this property would work
just as well. Thanks to FOTH Property \ref{transition} (which excludes
the possibility of FOTS that are partially isolated and partially
dynamical) we can always scale the null vectors so that 
\begin{equation}\label{dhB}
  \frac{B}{2G} =  \frac{dE}{dv} \, ,
\end{equation}
where $G$ is the gravitational constant.  This means that $B$ will be
constant on each leaf of the foliation\footnote{ This scaling is
equivalent to requiring that the pull-back of $\ell_a$ to $T^\star H$
satisfy $\underleftarrow{\ell} = - \frac{dB}{dv} dv$.}. With a bit of
rearranging, equation (\ref{Lvt}) integrates over $S_v$ to become
\begin{equation}
  \frac{dE}{dv} = \int_{S_v} \tilde{\epsilon} \left[ T_{ab} \ell^{a} \tau^{b}
  + \frac{A \norm \sigma_{(\ell)} \norm ^{2}}{8\pi G}
  + \frac{B \norm \tilde{\omega} \norm ^{2}}{8\pi G} \right] \, , \label
{dEodv}
\end{equation}
where we have used the fact that the $S_v$ is topologically $S^2$ to
rewrite $\int_{S_v} \vS \tilde{R} = 8 \pi$. Then, integrating up the
horizon we find that 
\begin{equation}
  E(v_2) - E(v_1) =  \int_{v_1}^{v_2} dv  
  \left\{ \int_{S_v} \tilde{\epsilon} \left[ T_{ab} \ell^{a} \tau^{b}
  + \frac{A \norm \sigma_{(\ell)} \norm^{2}}{8\pi G}
  + \frac{B \norm \tilde{\omega} \norm^{2}}{8\pi G} \right] \right\} 
  \, . \label{akFlux}
\end{equation}

If $E(v)$ is any state function of the horizon (such as energy or
entropy) this is interpreted as a flux law.  This flux law is valid for
\textit{any} functional $E(v)$, although Hayward \cite{haywardFlux} has
argued that the Hawking (or irreducible) mass is the most natural
choice.  Given the dominant  energy condition, the right-hand side of
(\ref{akFlux}) consists of three non-negative terms. The interpretation
of the first two is reasonably straightforward as a matter flux and a
flux of gravitational energy through the horizon.  However, the third
term is not so easily understood. A possible interpretation is that this
represents a flux of rotational energy \cite{ak, KrishNum, haywardFlux}
however this seems unlikely as $\tom_a$ is associated with the angular
momentum itself rather than its flux. 

A direct comparison in the slowly evolving limit with (\ref{FirstLaw})
cannot be made due to the different methods of scaling the null 
vectors.
The slowly evolving formalism can be generalized to allow for such a
comparison by relaxing the requirement that $A=1$,  however even 
then
the limit does not  go through directly.  To see this note that we can
expand the integrand of the right-hand side of the dynamical horizon
flux law as
\bea
  A \left( \norm \sigma^{(\ell)} \norm ^2/(8\pi G) 
  + T_{ab} \ell^a \ell^b  \right) 
  + B \left(\norm \tom \norm^2/(8 \pi G) 
  + T_{ab} \ell^a n^b \right) \, . \label{DynFlux}  
\eea
In order to make a comparison with the slowly evolving horizons, we
consider the limit where $A \approx 1$ and $B \approx \epsilon^{2}$.
Then, all of these terms are of order $\epsilon^2$ and the last two ---
which do not arise in the slowly evolving law --- cannot be neglected.
Further, for the apparently common terms, the $A \neq 1$ version of
the slowly evolving constraint law
(\ref{constraint_non_rot}) gives rise to shear and flux terms
\bea
  A^2\left( \norm \sigma^{(\ell)} \norm^2/(8\pi G) 
  + T_{ab} \ell^a \ell^b  \right) \, .
\eea
These differ by a factor of $A$ from those in (\ref{DynFlux}).

Clearly in the limit where the horizon is slowly evolving, this flux law
will reduce to (\ref{FirstLaw}) up to the factor of $A$ discrepancy
discussed above.  Now, it is quite possible that as a horizon 
approaches
equilibrium, $A$ will be constant at leading order over each
cross-section.  If this is the case, the two results will agree as we
can set $A = 1 + O(\epsilon)$.  However, there does not seem to be 
any
analytic justification for this, and the examples of slowly evolving
horizons studied so far in \cite{billsPaper} are all (approximately)
spherically symmetric, whence $A$ is automatically (approximately)
constant by construction. Thus, whether or not this assumption holds 
is
currently an open question.

Motivated by the results for slowly evolving horizons, we can derive an
alternative flux law which is valid on all dynamical horizons.  On a
dynamical horizon, we must have $\delta_{\ell} \theta_{(\ell)} \le 0$,
and $\delta_{n} \theta_{(\ell)} \le 0$ with strict inequality at some
point on the horizon.  If these conditions do not hold, then the horizon
will not be spacelike.  Making use of these conditions, we can rewrite
(\ref{Lvt}) as
\begin{equation}
  - B \delta_{n} \theta_{\ell} = - A \delta_{\ell} \theta_{(\ell)} \, .
\end{equation}
On integration over a cross section of the horizon, both sides are
guaranteed to be positive.  Therefore, making use of (\ref{dntl}) and
(\ref{dltl}) we obtain
\begin{equation}
   (1 - e) \left(\frac{B}{2G}\right)= 
  \int_{S} \frac{A \norm\sigma^{(\ell)}\norm^{2}}{8 \pi G} 
  +  A T_{ab} \ell^{a} \ell^{b}
\end{equation}
where
\begin{equation}
  e := \int_{S} \frac{1}{4\pi} \norm\tom^{2}\norm + 2G T_{ab} \ell^{a} n^
{b} 
\end{equation}
and $e$ is related to the extremality of the horizon as described in
detail in \cite{bfExtreme}.

Given the energy functional $E(v)$ we simply set 
\begin{equation}\label{altB}
  (1-e) \frac{B}{2G} = \frac{dE}{dv}
\end{equation}
(compare this with (\ref{dhB})).  Then given this normalization, the
dynamical horizon flux law becomes
\begin{equation}\label{altFlux}
  E(v_2) - E(v_1) =  \int_{v_1}^{v_2} dv  
  \left\{ \int_{S_v} \tilde{\epsilon} \left[ A T_{ab} \ell^{a} \ell^{b}
  + \frac{A \norm \sigma_{(\ell)} \norm^{2}}{8\pi G} \right] \right\}
  \, . 
\end{equation}
Although this flux law is similar to (\ref{dEodv}), there are obvious
differences.  Specifically, the $\norm \omega \norm^{2}$ and $T_{ab}
\ell^{a} n^{b}$ terms are no longer present --- they have been 
absorbed
into the definition of $B$, and consequently the scaling of the null
vectors.  In many ways, this is preferable as these terms appear to be
associated to intrinsic features of the horizon and \emph{not} to fluxes
of matter or gravitational energy through the horizon.  In particular,
for a charged black hole $e = Q^{2}/R^{2}$, while for the Kerr metric
$e$ is a function of $J/M^{2}$.

Despite the similar forms and origins of the dynamical and slowly
evolving flux laws, a closer analysis shows that they are actually quite
different.  We have managed to rewrite the dynamical horizon law in a
manner which more closely resembles the slowly evolving horizon 
result.
However, differences still remain. Further discussion of the dynamical
horizon flux law and its relation to other formalisms may be found in
\cite{ak, theBeast, KrishNum, meRev, GourgFlux}.

%
\section{Conclusions and Outlook}
%

By concentrating on the foliating two-surfaces that are the constituent
parts of future outer trapping horizons we have constructed a framework
that encompasses isolated, slowly evolving, and dynamical horizons.  The
techniques used are similar in spirit to previous geometric analyses of
horizons \cite{ams, Andersson:2005me}, although our focus here has been
on the deformations of MOTS in the full four-dimensional spacetime,
rather than a three-dimensional slice.  Furthermore, we have seen that
horizon evolutions and variations are both governed by the same
underlying equations. Additionally, this set of equations is responsible
for the various flux laws associated with these surfaces.  
Most of these results have been obtained previously by various
authors \cite{ams, hayward, haywardPert, haywardFlux, isoPRL, isoMech, ak, gregabhay, mttpaper, GourgRev} using a variety of methods. The contribution of this paper is to rederive
many of these properties of horizons using 
a common geometrical framework which highlights the connections
between the various results.

Thus we have seen that along with the freedom to refoliate an isolated
FOTH is the concomitant rigidity that prevents us from varying its
three-dimensional structure. In contrast, the uniqueness of the
foliation for a dynamical FOTH is the flip side of their well-known lack
of rigidity against out-of-surface deformations. The freedom to vary
dynamical FOTHs is also equivalent to fact that there is no unique
method for evolving a given FOTS into a FOTH. Instead its evolution is
many-fingered and any FOTS can be (locally) extended into many 
different
FOTHs. 

This provides some insight into one of the most interesting open
problems about these quasi-locally defined horizons. It is well-known
that apparent horizons (in this case defined as the outermost 
marginally
trapped surface on a spacelike slice $\Sigma$) can discontinuously 
jump
during especially dramatic events such as black hole mergers. 
However,
it is widely suspected (see for example \cite{mttpaper, BadApVar,
meRev}) that many of these discontinuities may result from the
interweaving of the foliation with a continuous $\tl = 0$ surface that
is sometimes a FOTH but at other times fails to satisfy one or both of
$\tn < 0$ and $\delta_n \tl < 0$. 

Therefore, it becomes important to understand the circumstances 
under
which a FOTH may end. We have seen that a FOTH may always be 
locally
extended however, this certainly does not guarantee the existence of
global extension. For example, in \cite{mttpaper} it was seen that on
horizons in Tolman-Bondi spacetimes, it is possible for $\delta_n \tl$
to switch signs and become positive.  At this point, the FOTH ends 
while
a future inner trapping horizon begins.  In these examples, there exists
a continuous three-surface foliated by $\theta_{(\ell)} = 0$
cross-sections, though only part of that surface can be identified as a
black hole horizon.  A question deserving further investigation is then:
what happens to FOTHs under finite extensions?  More specifically, 
can a
FOTH always be extended into an unending structure that is foliated 
by
$\tl = 0$ surfaces? These extensions could fail if, for example, the
repeated deformations broke the spacelike nature of the foliation
two-surfaces or if there are circumstances under which the extension 
is
open but bounded -- that is $H = \cup_v S_v$ with $v \in (v_1, v_2)$
where $v_1$ and/or $v_2$ are finite. 

On a related note we motivated our horizon definitions by the notion
that the interior of a black hole should consist of all points that lie
on a trapped surface.  However, we have seen that future outer 
trapping
horizons are typically only the boundary of the trapped region
associated to a particular slicing of the given spacetime. Another
important question is then: what is the real boundary of the trapped
region? It is widely believed \cite{gregabhay, BadApVar, meRev} that
this boundary corresponds to the event horizon in spacetimes where 
this
structure is defined, however this has not been proved. A study of the
finite extensions of FOTS where one tries to ``push" them towards the
event horizon would be one way of gaining insight into this problem. 

Returning to the results of this paper, a particularly interesting
application of the deformation rules comes in the detailed investigation
of the near-equilibrium slowly evolving horizons, originally introduced
in \cite{prl}. We characterized these as being almost-null in the sense
that the equations governing their evolutions are almost the same as
those for null surfaces in general and isolated horizons in particular.
Slowly evolving horizons were shown to obey a dynamical version of the
first law of black hole mechanics, just as in standard thermodynamics
near-equilibrium systems obey the $TdS$-form of the first law. This
result was first reported in \cite {prl} but the derivations appear here
for the first time. We also saw that this result follows from the same
``constraint" equations that are responsible for the corresponding
Hawking-Hartle formula for event horizons.  The fact that both of these
results depend on the horizons being almost non-expanding (as well as
some of the examples considered in \cite{billsPaper}) leads one to
speculate that an eternally slowly evolving horizon might be
indistinguishable from the corresponding event horizon to the same order
of accuracy for which the other results hold.  Further the rigidity of
isolated horizons against deformations suggests that slowly evolving
horizons might also be rigid up to this order. 

Our focus on the deformations of two-surface has been demonstrated 
to
unify and illuminate diverse results in the study of quasi-local
horizons.  It seems likely that it will continue to be useful in
studying future problems, including those outlined above. 
 
\section*{Acknowledgements}

We would like to thank Abhay Ashtekar, Christopher Beetle, Patrick
Brady, Jolien Creighton, Greg Galloway, Bill Kavanagh and Badri 
Krishnan
for helpful discussions.  I.B. was supported by the Natural
Sciences and Engineering Research Council of Canada.  S.F. was supported by 
NSF grant
PHY-0200852.

\begin{appendix}

%
\section{Deriving the two-surface equations}
\label{appA}
%

In this Appendix we catalogue some useful relations for two-surfaces
embedded in four-space and use them to derive the equations of 
section
\ref{twoSurf}. 

\subsection{Preliminaries}
\label{Prel} 
 
It is often useful to decompose the four-dimensional Riemann tensor
into its Weyl and Ricci tensor components. Thus we note that
\cite{hawkellis}:
\bea
\mathcal{R}_{abcd} = C_{abcd} 
+ \left( g_{a [ c} \mathcal{R}_{d] b} + g_{b [ d} \mathcal{R}_{c] a} \right) 
- \frac{1}{3} \left( g_{a [c} g_{d] b} \right) \mathcal{R} \, , \label{weyl} 
\eea
where $C_{abcd}$ is the Weyl tensor and as usual square brackets
indicate anti-symmetrization. 

Next, given a two-surface $(S,\tq_{ab}, d_a)$ in four-space $(M, g_
{ab},
\nabla_a)$ the Gauss, Codazzi, and Ricci equations relate the 
curvature
of spacetime to the intrinsic and extrinsic geometry of the two-surface.
They may be derived fairly directly from a few facts.  First, the
push-forward of the inverse two-metric on $S$ can be written as 
\bea
\tq^{ab} = g^{ab} + \ell^a n^b + \ell^b n^a \, , 
\eea 
for any two null normals to $S$ that satisfy $\ell^a n_a = -1$. Second,
covariant derivatives of tensors defined intrinsically to $S$ may be
written in terms of the full four-dimensional derivative with the help
of appropriate projections. Thus, for example, for a one-form $\mu_a 
\in
T^{\star} S$, 
\bea
d_a d_b \mu_c = \tq_a^d \tq_b^e \tq_c^f \nabla_d 
(\tq_e^g \tq_f^h \nabla_g \mu_h ) \, . 
\label{TwoDer} 
\eea
Finally, the Riemann tensors on $M$ and $S$ respectively satisfy
\bea
(\nabla_a \nabla_b - \nabla_b \nabla_a) W_c 
&=& \mathcal{R}_{abcd} W^d \, \, \mbox{and} \\
(d_a d_b - d_b d_a) w_c &=& \tilde{R}_{abcd} w^d \, , \label{R2}
\eea
for one-forms $W_a \in T^\star M$ and  $w_a \in T^\star S$. 

Then substituting (\ref{R2}) into (\ref{TwoDer}) and doing some 
algebra,
one can show that 
\bea
\tq_a^e \tq_b^f \tq_c^g \tq_d^h \mathcal{R}_{efgh} 
= \tilde{R}_{abcd}
+ (k^{(\ell)}_{ ac}  k^{(n)}_{ bd} + k^{(n)}_{ ac}  k^{(\ell)}_{ bd}) 
-  (k^{(\ell)}_{ bc}  k^{(n)}_{ ad} + k^{(n)}_{ bc}  k^{(\ell)}_{ ad}) \, \, , 
\label{Gauss}
\eea
where $k^{(\ell)}_{ab}$ and $k^{(n)}_{ab}$ are the extrinsic curvatures
defined in (\ref{kln}).  This is the Gauss relation. 

Through similar calculations one can also derive the Codazzi relations: 
\bea
\tq_a^e \tq_b^f \tq_c^g \ell^h \mathcal{R}_{efgh} 
&=& (d_a - \tom_a) k^{(\ell)}_{ bc} 
- (d_b - \tom_b) k^{(\ell)}_{ ac} \label{Codazzi} \\
\tq_a^e \tq_b^f \tq_c^g n^h \mathcal{R}_{efgh} 
&=& (d_a + \tom_a) k^{(n)}_{ bc} 
- (d_b + \tom_b) k^{(n)}_{ ac} \nn
\eea
where $\tom_a$ is the normal bundle connection defined in equation
(\ref{tom}). Alternatively, applying (\ref{dq}) and (\ref{weyl}) these
may be expanded as 
\begin{eqnarray}
  ({d}_{a} - \tom_a) \theta_{(\ell)} &=&
  2 ({d}_{b} - \tom_b) \sigma^{(\ell) b}_a
  - \tilde{q}_{a}^{b}  G_{bc}\ell^{c} 
    - 2 \tq_a^b  C_{bcde} \ell^{c} \ell^{d} n^{e} \mbox{ and}
  \label{da_thetal}\\
  ({d}_{a} + \tom_a) \theta_{(n)} &=& 
  2 ({d}_{b} + \tom_b) \sigma^{(n)  b}_a 
  - \tilde{q}_{a}^{b} G_{bc} n^{c} 
    + 2 \tq_a^b C_{bcde} n^{c} \ell^{d} n^{e}  \, \, . 
\end{eqnarray}
Finally, by the same kinds of calculations used to derived the Gauss 
and
Codazzi relations we can also derive the Ricci relation
\bea
\tq_a^c \tq_b^d \ell^e n^f C_{cdef} = 
\tq_a^c \tq_b^d \ell^e n^f \mathcal{R}_{cdef} 
=  \Omega_{ab} - \sigma^{(\l)}_{a  c} \sigma^{(n) c}_b 
+  \sigma^{(\ell)}_{ b  c} \sigma^{(n) c}_a 
\label{Ricci}
\eea
where $\Omega_{ab}$ is the curvature of the normal bundle defined in
equation (\ref{Om}).

\subsection{Deformation equations}
\label{DefEq}

Next we consider the derivation of the deformation equations. As an
initial step towards calculating $\delta_X \tl$ we find $\delta_X
k^{(\ell)}_{ab}$.  To this end a few carefully chosen lines of algebra
along with an application of equation (\ref{LXl}) show that:
\bea
\delta_X k^{(\ell)}_{ ab} = 
  - \frac{1}{2} \tq_a^c \tq_b^e (X^d \ell^f + X^f \ell^d) \mathcal{R}_{cdef} 
  - \ell^f \tq_{(a}^c \tq_{b)}^d \nabla_c (\tq_d^e \nabla_e X_f) 
  + \kappa_X k^{(\ell)}_{ ab} \, , 
\eea
where the round index brackets indicate the usual symmetrization of 
the
enclosed indices.  Now equations (\ref{weyl}) and (\ref{Gauss}) can be
used to rewrite the first term, while the second can be shown to be 
\bea
- \ell^f \tq_a^c \tq_b^d \nabla_c (\tq_d^e \nabla_e X_f)) = 
k^{(\ell)}_{ ac} k^{(X) c}_b - d_a d_b B
+ B d_a \tom_b +  2\tom_{(a} d_{b)} B  - B \tom_a \tom_b \, \, . 
\eea
Then, we find that 
\bea
\delta_X k^{(\ell)}_{ab} & = & 
  - d_a d_b B + 2 \tom_{(a} d_{b)} B 
  + \kappa_X k^{(\ell)}_{ ab} \\
&& +  A \left(k^{(\ell)}_{ac} k^{(\l) c}_b 
  - \tq_a^c \l^d \tq_c^e \l^f C_{cdef} 
  - \frac{1}{2} \tq_{ab} G_{cd} \l^c \l^d \right) \nn\\
&& + B \left( \frac{1}{2} \tilde{R}_{ab} 
  + \frac{1}{2} [ \tl k^{(n)}_{ab} + \tn k^{(\ell)}_{ab}]  
  -  2 k^{(\ell)}_{c(a} k^{(n) c}_{b)}  \right) \nn \\
& & + B \left(- \frac{1}{2} \tq_a^c \tq_b^d \mathcal{R}_{cd} 
  + d_{(a} \tom_{b)} - \tom_a \tom_b \right) 
\, \, . \nn
\eea
With the help of (\ref{Lq}) it is then straightforward to show that
\begin{eqnarray} 
   \delta_X \theta_{(\ell)} - \kappa_X \tl &=& 
   - {d}^{\, 2} \mspace{-2mu} B  
   + 2 \tilde{\omega}^{a} {d}_{a} B 
   - B \left[ \norm\tilde{\omega}\norm^2
     -  {d}_{a} \tilde{\omega}^{a} 
     - \tilde{R} /2
     + G_{ab} \ell^{a} n^{b} - \tl \tn \right] \nn \\
     & &    \quad    - A \left[ \norm\sigma^{(\ell)}\norm^{2} 
     + G_{ab} \ell^{a} \ell^{b}
     + (1/2) \theta_{(\ell)}^{2} \right]   \, , 
\end{eqnarray}
where $\norm \tom \norm^2 = \tom_a \tom^a$ and  $\norm \sigma^
{(\ell)}
\norm^2 = \sigma^{(\ell)}_{ab} \sigma^{(\l)\, ab}$.  Similarly, we can
take the trace-free part of the above to obtain
\bea
\delta_X \sigma^{(\ell)}_{ ab} - \kappa_X \sigma^{(\ell)}_{ ab} & = &
- {d}_{\{ a} {d}_{b\} } B + 2 \tom_{ \{ a} {d}_{b \} } B  
- A \tq_a^c \ell^d \tq_b^e \ell^f C_{cdef} \label{siglVar} \\
& & +  B \left[\frac{1}{2} \tl \sigma^{(n)}_{ab} 
  - \frac{1}{2} \tn \sigma^{(\l)}_{ ab}
  + {d}_{\{ a} \tom_{b \}} - \tom_{ \{ a} \tom_{b \} } 
  - \frac{1}{2}  \tq_{\{ a}^c \tq_{b \}}^d G_{cd} \right] \nn \\
& &  + A \left[ \sigma^{(\l)}_{ac} \sigma^{(\l) c}_{b}  
  + \frac{1}{2} \norm\sigma^{(\ell)} \norm^2 \tq_{ab}  \right] 
  -  B \left[ \sigma^{(\l) }_{ac} \sigma^{(n) c}_{b}  
  +  \sigma^{(\l)}_{bc} \sigma^{(n) c}_{a} \right]  \nn \, . 
\eea
Here curly brackets around a pair of indices indicates their
symmetric trace-free part (with respect to the two-metric). Thus, for
example, 
\bea
\tom_{ \{ a  } {d}_{b \} } B = \tom_{ ( a  } {d}_{b ) } B
  - \frac{1}{4} \tq_{ab} 
  \tom^c {d}_c B \, . 
\eea
Note that even though $\sigma^{(\ell)}_{ ab}$ is trace-free, its
variation will not usually inherit this property (this follows from the
fact that $\delta_X \tq^{ab} \neq 0$). Thus, in the above expression,
the first two lines are trace-free while the last line is not. 

A direct expansion of $\tq_a^b \Lie_X (n_c \nabla_b \ell^c)$
with applications of (\ref{LXl}) and (\ref{weyl}) gives us the variation
of the angular momentum one-form:
\begin{eqnarray}
    \delta_X \tilde{\omega}_{a}  - {d}_a \kappa_X  &=& 
   -
   k^{(\ell)}_{ a b}
   \left[ {d}^b A  +  \tilde{\omega}^b A \right]
  - k^{(n)}_{ab}
  \left[{d}^b B - \tilde{\omega}^b B 
    \right] \\
 & &   + \tilde{q}_{a}^{\; \; b} \left[ \frac{1}{2} G_{bc}\tau^{c}
    - C_{bcde} {X}^{c}\ell^{d}n^{e} \right]    \nonumber \, .
\end{eqnarray}
Finally, we can calculate the deformation of the two-curvature.  Taking
the standard variation of the Ricci scalar (which is used, for example,
in deriving the Einstein equations from the Einstein-Hilbert action
\cite{hawkellis}) and adapting it to two-dimensions, the variation  
\bea
\delta ( \vS \tilde{R}) = 
  \vS (\tilde{R}_{ab} - 1/2 \tilde{R} \tq_{ab} ) \delta \tq^{ab}  
  + \vS d_a \left( \tq^{ad} \tq^{bc} 
  [ d_c \delta \tq_{bd} - d_d \delta \tq_{bc} ] \right) \, \, . 
\eea
Taking the $\delta$s as $\delta_X$s and doing a little algebra this
becomes
\bea
\delta_X (\vS \tilde{R}) = 
  2  \vS d^a d^b (A  \sigma^{(\ell)}_{ab} - B \sigma^{(n)}_{ab} ) 
  - \vS d^2(A \tl - B \tn) \, , 
\eea
since the $\tilde{R}_{ab} = \frac{1}{2} \tilde{R} \tq_{ab}$ in two
dimensions. Then, 
\begin{equation}\label{2curv_var}
  \delta_X \tilde{R} = 
  2  d^a d^b (A  \sigma^{(\ell)}_{ab} - B \sigma^{(n)}_{ab} ) 
  - d^2 (A \tl - B \tn)  - (A \tl - B \tn) \tilde{R} \, . 
\end{equation}

%
\section{Maximum and minimum principles on $S^2$}
\label{maxmin}
%

In this Appendix we briefly review the maximum principle for linear
second order elliptic partial differential operators and then apply it
to operators and functions defined on a surface which is diffeomorphic
to $S^2$. 

As motivation we begin with a local maximum principle. 
Let $U \subset \Rbar^n$ be an open set parameterized by coordinates
$x^i$ where $i \in \{1,2, \dots n\}$. 
Then any second order differential operator on the
set of twice-differentiable functions over $U$ takes the form
\bea
L[f] = \alpha^{ij} \frac{\partial^2 f}{\partial x^i \partial x^j} 
+ \beta^i \frac{\partial f}{\partial x^i} 
+ \gamma f \, , 
\label{Lf}
\eea
for some functions $\alpha^{ij}$, $\beta^i$, and $\gamma$ where $i, j$
are indices and we assume the usual summation convention. If
$\alpha^{ij}$ is positive definite then we say that $L$ is elliptic. 

Now if $f$ satisfies $L[f] > 0$ everywhere on $U$ for an elliptic $L$
with $\gamma \leq 0$, it is easy to show that $f$ cannot have a
non-negative local maximum in $U$. To see this recall from 
elementary
calculus that if $f$ has a local maximum at $p$, then $\partial f /
\partial x^i= 0$ for all $i$ and the matrix $\partial^2 f / \partial x^i
\partial x^j$ is negative definite. Equally elementary linear
algebra tells us that for positive definite $\alpha^{ij}$ we must have
\bea
\alpha^{ij} \frac{\partial^2 f}{\partial x^i \partial x^j} < 0 \, \,  
\eea
at $p$. Then a term-by-term analysis of the right-hand side of
(\ref{Lf}) quickly shows that if $f\geq0$ at $p$ then $L[f] < 0$  in
contradiction to the original assumption.  Therefore, if a local maximum
exists, it must be negative. 

Note that as formulated this result doesn't cover cases where 
$\mbox{det} (\partial^2 f / \partial x^i \partial x^j) = 0$ at $f_{max}$.
However, the principle may be extended to all maxima and
that is the content of the following theorem which is stated and proved
in, for example, \cite{spiv5}. 

\newtheorem{thm}{Theorem}
\begin{thm}[Hopf's Maximum Principle] \label{hopf} 
Consider a second order differential operator 
\bea
L[f] = \alpha^{ij} \frac{\partial^2 f}{\partial x^i \partial x^j} 
+ \beta^i \frac{\partial f}{\partial x^i} 
+ \gamma f \, , 
\eea
on a connected open set $U \in \Rbar^n$ over which $\gamma \leq 0$.
Assume that the functions $\beta^i$ and $\gamma$ are locally 
bounded and
that in a neighbourhood of any point of $U$ there are constants $M_1
$
and $M_2$  such that 
\bea
M_1 (\xi \cdot \xi) \leq \alpha^{ij} \xi_i \xi_j \leq M_2 (\xi \cdot \xi) \, , 
\eea
for all $\xi_i \in \Rbar^n$, where $\cdot$ is the usual Euclidean dot
product for $\Rbar^n$ and  we understand summation over repeated
indices. Finally let $f \in C^2 (U)$ and assume that it satisfies
\bea
L[f] \geq 0 \, , 
\eea 
everywhere in $U$. Then $f$ cannot have a non-negative maximum 
on $U$,
unless $f$ is everywhere constant. 
\end{thm}

\newtheorem{cor}{Corollary}

We can then apply this theorem to show the following: 

\begin{cor}[Maximum principle on a two-sphere]
Let $S$ be a two-manifold that is topologically $S^2$ and has 
spacelike
two-metric $\tq_{ab}$ and compatible covariant derivative $d_a$.
Further, let $f \in C^2(S)$ be a scalar field that satisfies
\bea
L[f] \geq 0 \, \, , 
\eea
everywhere on $S$ for a differential operator of the form 
\bea
L[f] = \tq^{ab} d_a d_b f + \beta^a d_a f + \gamma f \, \, , 
\eea
where $\beta^a \in TS$ and $\gamma \leq 0$. Then $f$ is either 
constant
or everywhere negative.  \end{cor}

\noindent To see this first note that $S^2$ is compact and so $f$ must
have (and achieve) an absolute maximum.  However, if we consider 
any
finite set of charts that cover $S$, Theorem \ref{hopf} applies to the
coordinate realization of $L[f]$ on each of those charts. Thus working
chart-by-chart and piecing the results together, either $f$ is constant
over $S$ or the absolute maximum $f_{max} < 0$.  That is, $f$ is 
either
everywhere constant or everywhere negative. 
$\Box$

Similarly with a simple substitution $f \rightarrow -f$ we have a
minimum principle:

\begin{cor}[Minimum principle on a two-sphere]
Let $S$ be a two-manifold that is topologically $S^2$ and has 
spacelike
two-metric $\tq_{ab}$ and compatible covariant derivative $d_a$.
Further, let $f \in C^2(S)$ be a scalar field that satisfies
\bea
L[f] \geq 0 \, \, , 
\eea
everywhere on $S$ for a differential operator of the form 
\bea
L[f] = - \tq^{ab} d_a d_b f + \beta^a d_a f + \gamma f \, \, , 
\eea 
where $\beta^a \in TS$ and $\gamma \geq 0$. Then $f$ is either 
constant
or everywhere positive. 
\end{cor}

These results are used repeatedly in section \ref{FOTS}. 

%
\section{Classification of Kerr horizons}
\label{KerrFOTS}
%

To justify some of the assumptions made in defining FOTHs and 
slowly
evolving horizons, we consider the values that some of the key 
geometric
quantities take in the Kerr spacetime\footnote{Similar results hold for
the Kerr-Newmann family, but for ease of presentation we restrict our
attention to pure Kerr. }.  This should certainly contain both a FOTH
and a slowly evolving horizon (with $\epsilon = 0$) so we must verify
that the various assumption hold on the Kerr horizon. Further, for
spacetimes that are perturbations of Kerr the values of these 
quantities
should be similar and so these calculations gives us some intuition 
about
these spacetimes and their classification as well. 

In generalized ingoing Eddington-Finkelstein coordinates, the Kerr 
metric takes the
form
\bea
ds^2 &=& - \left(1- \frac{2mr}{\Sigma} \right) dv^2 + 2 dv dr 
- \frac{4amr \sin^2 \theta}{\Sigma} dv d\phi 
+ \Sigma d\theta^2 \\
& & + \frac{\sin^2 \theta \left((r^2+a^2)^2 
- a^2 \sin^2 \theta (r^2-2mr+a^2) \right)}{\Sigma} d \phi^2 \, , 
\nn
\eea
where $\Sigma = r^2 + a^2 \cos^2 \theta$ and the event horizon is at 
$r
= m + \sqrt{m^2-a^2}$. 

\begin{figure}
\begin{center}
\scalebox{.9}{\includegraphics{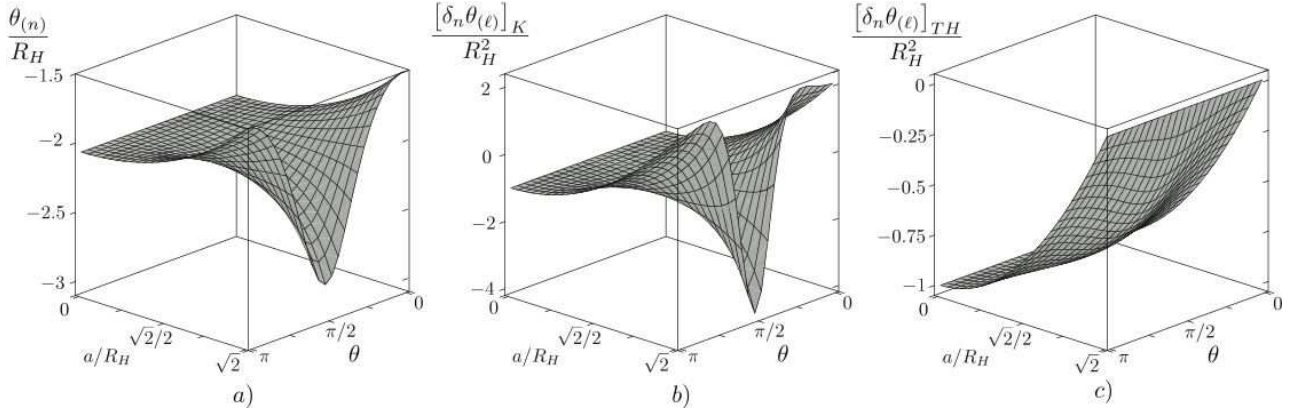}}
\end{center}
\caption{The FOTS defining quantities for a Kerr horizon. That $\tn < 0
$
everywhere on all Kerr horizons is shown by a) which graphs it for the
standard scaling of the null vectors.  However b) and c) demonstrate
that the $\delta_n \tl < 0$ condition is not so trivial. For the
standard (\ref{lK}) scaling of the null vectors $\delta_n \tl$ can be
positive and we must rescale as in equation (\ref{lTH}) to show that the
condition is always satisfied.}
\label{KerrFig_FOTS}
\end{figure}

It is common to scale the null vectors so that $\ell^a$ is proportional
to the global ``time translation" Killing vector field on $H$ and the
corresponding flow evolves the $v=\mbox{constant}$ two-surfaces into
each other. Identifying this scaling with a sub/superscript-``$K$" we
have 
\bea
\ell_{K}^a &=& \left[1, 0, 0, \frac{a}{r^2+a^2} \right] \, \, 
\mbox{and} \label{lK}
\\
n^{K}_{a} &=&  \left[ - 1, \frac{a^2 \sin^2 \theta}{2(r^2+a^2)},0,0 \right]
\nn \, \, , 
\eea
where the coordinate ordering is $\{ v,r,\theta,\phi \}$. This is also 
the scaling for which the surface gravity is constant and takes the 
value
\bea
\kappa_\ell = \frac{1}{2r} \left(\frac{r^2-a^2}{r^2+a^2}\right) \, . 
\eea 

In Figures \ref{KerrFig_FOTS}a and \ref{KerrFig_FOTS}b we plot $\tn$ 
and
$\delta_n \tl$ respectively  for this scaling of the null vectors. It is
immediately apparent that while $\tn < 0$, horizons with sufficiently
large angular momentum have $\delta_n \tl \geq 0$ over some region.
However, this does not imply that for these horizons the $v =
\mbox{constant}$ surfaces fail to be FOTHs.  Instead it nicely
demonstrates that different scalings of the null vectors generate
different deformations of the two-surfaces and that not all of the
resulting two-surfaces are fully trapped. If we rescale the null vectors
to become 
\bea
\ell_{TH}^a = \frac{1}{\Sigma} \ell_{K}^a
\, \, \mbox{and} \, \, 
n^{TH}_{a} = \Sigma n^{K}_{a} \,  , \label{lTH}
\eea
then $\delta_n \tl$ calculated with respect to these vectors is
everywhere negative, as shown in Figure \ref{KerrFig_FOTS}c. Thus 
the $v
= \mbox{constant}$ slices are FOTS, but the standard scaling of the 
Kerr
null normals does not show this. 

\begin{figure}
\begin{center}
\scalebox{.9}{\includegraphics{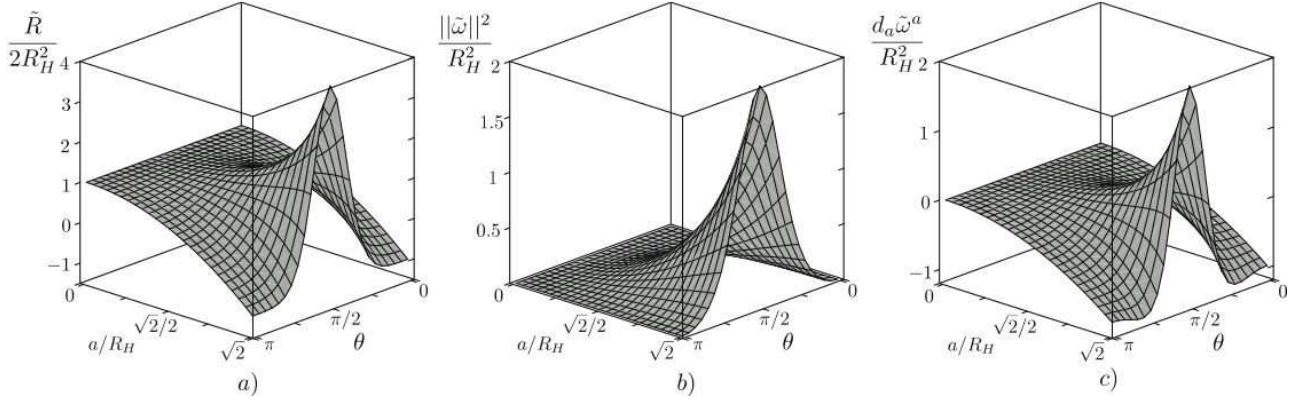}}
\end{center}
\caption{The quantities that determine $\delta_n \tl$ (which also 
appear 
in the first condition of the definition of a slowly expanding horizon) 
as calculated for the standard scaling of the null vectors (\ref{lK}).}
\label{KerrFig_dntlComp}
\end{figure}

Computationally we can understand what is happening for these two
different scalings by considering the individual components that went
into the calculations. Equation (\ref{dntl}) tells us that in the
absence of matter fields
\bea
\delta_n \tl = - \tilde{R}/2 + \norm\tom\norm^2 - d_a \tom^a \, . 
\eea 
Figure \ref{KerrFig_dntlComp} graphs these quantities for the standard
scaling (\ref{lK}) of the null vectors.  The Ricci scalar is, of course,
scaling invariant and Figure \ref{KerrFig_dntlComp}a shows that for
horizons with large angular momentum it can be negative. Thus, the
scaling dependent $d_{a} \tom^{a}$ term must be positive and large
enough to compensate for this.  However, a quick examination of the
shapes of Figures \ref {KerrFig_dntlComp}b and \ref
{KerrFig_dntlComp}c
shows that for the standard rescaling, this term is also negative.

In fact these figures are also sufficient to show why the rescaling
(\ref{lTH}) resolves this problem --- although in a slightly roundabout
way. To see this we first note that for any (topologically) spherical
surface embedded in a  spacetime, there is always a scaling of the null
vectors so that $d_a \tom^a = 0$ (as discussed in \cite{JK, isoGeom}
this ultimately follows from the Hodge decomposition theorem). For the
two-surface under consideration this scaling is 
\bea
\ell_o^a = \frac{1}{\sqrt{\Sigma}} \ell_K^a 
\, \, \mbox{and} \, \, 
n^o_a = \sqrt{\Sigma} n^K_a \, . 
\eea
Taking this as a reference, equation (\ref{tom}) tells us that if
$\ell^a = f \ell_o^a$ and $n_a = (1/f) n_a^o$ then 
\bea
\norm \tom \norm^2 &=& \norm\tom_o\norm^2+ 2 \tom_o^a d_a
\ln f + \norm d \ln f\norm^{2} \, \, \mathrm{and} \label{tomf} \\ 
d_a \tom^a &=& d^2 \ln f \, \, . 
\eea
On an axisymmetric horizon, the second term on the right-hand side of
(\ref{tomf}) will vanish.  Meanwhile, $\norm \tom \norm^{2}$ is
invariant under a rescaling $f \rightarrow 1/f$, while $d_{a} \tom^{a}
\rightarrow -d_{a} \tom^{a}$. 
Just such a rescaling is made in the change from the Killing 
normalized
null vectors (\ref{lK}) to the FOTH normalized ones (\ref{lTH}), where
in this case $f = 1/\sqrt{\Sigma}$. Thus, it follows that
\bea
\left[ \delta_n \tl \right]_{TH} =  
- \tilde{R}/2 + \norm\tom_K\norm^2 + d_a \tom_K^a \, , 
\eea
and in this case the $d_{a} \tom^{a}$ term is preciesly what is required
to guarantee that $\delta_n \tl > 0$ over the whole horizon. 

Figure \ref{KerrFig_dntlComp} also demonstrates that $\tilde{R}$, $
\norm
\tom \norm^2$, and $d_a \tom^a$ are each of order $1/R_H^2$ as 
required
for a slowly expanding horizon.  Combining this with the fact that the
expansion parameter $C=0$ for the $v = \mbox{constant}$ two-
surfaces, we
confirm that the event horizon of a Kerr black hole is not only a FOTH
but is also a slowly evolving horizon. 

\begin{figure}
\begin{center}
\scalebox{.9}{\includegraphics{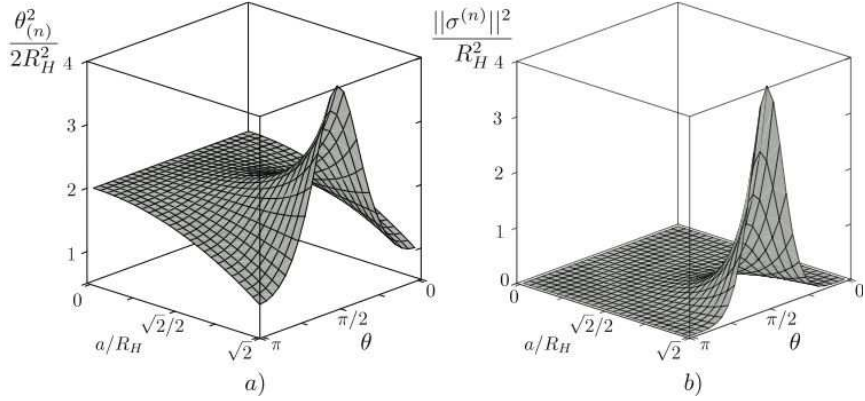}}
\end{center}
\caption{The quantities appearing in the definition of the slowly
evolving parameter $\epsilon$ as calculated for the standard scaling of
the null vectors (\ref{lK}).}
\label{KerrFig_eps}
\end{figure}

Finally, Figure \ref{KerrFig_eps} graphs the quantities that appear in
(\ref{epsilon}) in the definition of a slowly expanding horizon.
Although in this case their exact form doesn't matter (since
$C=\epsilon=0$), it is useful to note that they too are of order
$1/R_H^2$. Thus, for perturbed Kerr solutions (\ref{epsilon}) will be
satisfied and the horizon will be slowly expanding, as long as $C$ is
sufficiently small. 

\end{appendix}


\begin{thebibliography}{99}

\bibitem{hawkellis} S.~W.~Hawking and G.~F.~R.~Ellis, \textit{The 
Large
Scale Structure Sf spacetime} (Cambridge University Press, 1973). 

\bibitem{penrose} R.~Penrose, Phys.~Rev.~Lett. \textbf{14} 57 (1965).  

\bibitem{eardley} D.~M.~Eardley  Phys.~Rev.~D {\bf 57}  2299 (1998).

\bibitem{waldiyer} R.~M.~Wald and V.~Iyer, Phys.~Rev.~D {\bf 44} 
R3719 (1991).

\bibitem{BadApVar} E.~Schnetter and B.~Krishnan Phys.~Rev.~D {\bf 
73} 021502(R) (2006).

\bibitem{thomas} T.~W.~Baumgarte and S.~L.~Shapiro, 
Phys.~Reports {\bf 376} 41 (2003). 

\bibitem{hayward} S.~A.~Hayward, Phys.~Rev. D {\bf49} 6467 (1994).

\bibitem{haywardPert} S.~A.~Hayward, Phys.~Rev.~D {\bf 61}  
101503
(2000); Phys.~Rev.~D {\bf 64}  044002 (2001); Class.~Quant.~Grav. 
{\bf
18}  5561 (2001); Class.~Quant.~Grav. {\bf 23}  L15 (2006).

\bibitem{haywardFlux} S.~A.~Hayward, Phys.~Rev.~Lett. {\bf 93}  
251101
(2004); Phys.~Rev.~D {\bf 70} 104027 (2004); ``Conservation Laws for
Dynamical Black Holes", gr-qc/0607081. 

\bibitem{isoPRL} A.~Ashtekar, C.~Beetle, O.~Dreyer, S.~Fairhurst,
B.~Krishnan, J.~Lewandowski, J.~Wisniewski, Phys.~Rev.~Lett. {\bf 
85}
3564 (2000).

\bibitem{isoMech}  A.~Ashtekar, S.~Fairhurst and B.~Krishnan, 
\textit{Phys.Rev.} \textbf{D62} 104025 (2000);
 A.~Ashtekar, C.~Beetle and J.~Lewandowski, 
\textit{Phys.Rev.} \textbf{D64} 044016 (2001).
 
\bibitem{isoGeom} J.~Lewandowski and T.~Pawlowski, 
Int.~J.~Mod.~Phys.~D
{\bf 11}  739 (2002); A.~Ashtekar, C.~Beetle, and J.~Lewandowski,
Class.~Quant.~Grav. {\bf 19} 1195 (2002); J.~Lewandowski and
T.~Pawlowski, Class.~Quant.~Grav. {\bf 20} 587 (2003); 
J.~Lewandowski
and T.~Pawlowski, ``Symmetric non-expanding horizons", gr-qc/
0605026
(2006).

\bibitem{BadIsoNum}  O.~Dreyer, B.~Krishnan, E.~Schnetter and
D.~Shoemaker, Phys.~Rev.~D {\bf 67} 024018  (2003).  

\bibitem{LewIso} T.~Pawlowski, J.~Lewandowski and J.~Jezierski,
\textit{Class.~Quant.~Grav.} \textbf{21} 1237 (2004). 

\bibitem{ak} A.~Ashtekar and B.~Krishnan,
\textit{Phys.Rev.Lett.} \textbf{89}  261101 (2002); \textit{Phys.~Rev.}
\textbf{D68} 104030 (2003). 

\bibitem{gregabhay} A.~Ashtekar and G.~Galloway,
\textit{Adv.~Theor.~Math.~Phys.} \textbf{9} 1 (2005).

\bibitem{theBeast}  I.~Booth and S.~Fairhurst,
\textit{Class.~Quant.~Grav.} \textbf{22} 4515 (2005). 

\bibitem{ams} L.~Andersson, M.~Mars, and W.~Simon, 
{Phys.~Rev.~Lett.}
\textbf{95} 111102 (2005) .


\bibitem{KrishNum} E.~Schnetter, B.~Krishnan and F.~Beyer 
Phys.~Rev.~D
{\bf 74}  024028 (2006).

\bibitem{mttpaper} I.~Booth, L.~Brits, J.~A.~Gonzalez and C.~Van 
Den
Broeck,  Class.~Quant.~Grav. {\bf 23}  413 (2006).

\bibitem{bart} R.~Bartnik and J.~Isenberg, Class.~Quant.~Grav. {\bf 
23}
2559 (2006).

\bibitem{matt} A.~Nielsen and M.~Visser, Class.~Quant.~Grav. {\bf 23}
4637 (2006).

\bibitem{GourgFlux} E.~Gourgoulhon, Phys.~Rev.~D \textbf{72}, 
104007 (2005); 

\bibitem{Gourgoulhon:2006uc}
  E.~Gourgoulhon and J.~L.~Jaramillo,
  Phys.\ Rev.\  D {\bf 74}, 087502 (2006).

\bibitem{MK} M.~Korzynski, ``Isolated and dynamical horizons from a
common perspective",  gr-qc/ 0605019. 

\bibitem{prl} I.~Booth and S.~Fairhurst, \textit{Phys.~Rev.~Lett.}
\textbf{92} 011102 (2004). 

\bibitem{billsPaper} W.~Kavanagh and I.~Booth, Phys.~Rev.~D \textbf
{74}
044027 (2006). 

\bibitem{GourgRev} E.~Gourgoulhon and J.~L.~Jaramillo , 
Phys.~Rep.
\textbf{423}, 159 (2006). 

\bibitem{akRev} A.~Ashtekar and B.~Krishnan, Liv.~Rev.~Relativity 
{\bf
7}, 10 (2004). 

\bibitem{meRev} I.~Booth, Can.~J.~Phys. {\bf 83} 1073 (2005). 

\bibitem{Andersson:2005me}
  L.~Andersson and J.~Metzger,
  ``Curvature estimates for stable marginally trapped surfaces,''
  arXiv:gr-qc/0512106.

\bibitem{by} J.~Brown and J.W.~York, {Phys.~Rev.~D}  \textbf{47} 
1407
(1993).

\bibitem{HaywardDualNull} S.~A.~Hayward, Class.~Quant.~Grav.~
\textbf{10}, 779 (1993).

\bibitem{newman} R.~Newman, \textit{Class.~Quant.~Grav.} {\bf 4} 
277
(1987). 

\bibitem{pp} J.~M.~M.~Senovilla,  \textit{JHEP},  \textbf{11} 046
(2003). 

\bibitem{bfExtreme} I.~Booth and S.~Fairhurst, ``Extremality 
conditions
for dynamical horizons" (in preparation). 

\bibitem{hawkhartle} S.~W.~Hawking and J.~B.~Hartle, 
Commun.~Math.~Phys.
{\bf 27} 283 (1972).

\bibitem{fatherOfTheBeast} I.~Booth and S.~Fairhurst,
Class.~Quant.~Grav. {\bf 20}  4507 (2003). 




\bibitem{haywardRef}
  S.~A.~Hayward,
  Phys.\ Rev.\  D {\bf 74}, 104013 (2006);
  Class.\ Quant.\ Grav.\  {\bf 24}, 923 (2007).

\bibitem{spiv5} M.~Spivak, \textit{A comprehensive introduction to
differential geometry : Volume 5, Third Edition} (Publish or Perish,
Inc.\ : Houston) (1999).  

\bibitem{JK} J.~Jezierski, J.~Kijowski and E.~Czuchry, 
Rep.~Math.~Phys.
\textbf{46} 399-418 (2000).

\end{thebibliography}
\end{document}